\documentclass[12pt, draftclsnofoot, onecolumn]{IEEEtran} 

\usepackage{amsmath}
\usepackage{epsfig}
\usepackage{eucal}
\usepackage{amssymb}
\usepackage{bm}
\usepackage{cite}
\usepackage{graphicx}
\usepackage{multirow}
\usepackage{subfigure}
\usepackage{graphicx,color,psfrag}
\usepackage{multirow}
\usepackage{geometry}

\usepackage{array, tabularx, colortbl, color, threeparttable}
\usepackage[table]{xcolor}
\usepackage{psfrag, pstricks}
\usepackage{booktabs}

\newtheorem{lem}{Lemma}
\newtheorem{thm}{Theorem}
\newtheorem{cor}{Corollary}
\newtheorem{Def}{Definition}

\newcommand{\SINR}{\mathsf{SINR}}
\newcommand{\SNR}{\mathsf{SNR}}

\newcommand{\PG}{\mathsf{PG}}

\newcommand{\ASE}{\mathsf{ASE}}

\newcommand{\paren}[1]{\left(#1\right)}
\newcommand{\sqparen}[1]{\left[#1\right]}
\newcommand{\brparen}[1]{\left\{#1\right\}}
\newcommand{\abs}[1]{\left| #1\right|}
\newcommand{\field}[1]{\ensuremath{\mathbb{#1}}}
\newcommand{\R}{\ensuremath{\field{R}}} 
\newcommand{\Rp}{\ensuremath{\R_+}} 
\newcommand{\I}[1]{\ensuremath{\mathsf{1}_{\left\{#1\right\}}}} 
\newcommand{\PR}[1]{\ensuremath{\mathsf{Pr}\left\{#1\right\}}} 
\newcommand{\PRP}[1]{\ensuremath{\mathsf{Pr}\left(#1\right)}} 
\newcommand{\EW}{\ensuremath{\mathsf{E}}} 
\newcommand{\ES}[1]{\ensuremath{\mathsf{E}\left[#1 \right]}} 
\newcommand{\V}[1]{\ensuremath{\mathsf{Var}\left(#1 \right)}} 
\newcommand{\VS}[1]{\ensuremath{\mathsf{Var}\left[#1 \right]}} 
\newcommand{\defeq}{\ensuremath{\triangleq}} 
\newcommand{\e}[1]{\ensuremath{{\rm e}^{#1}}} 

\newcommand{\sinr}{\ensuremath{{\sf SINR}}}
\newcommand{\sir}{\ensuremath{{\sf SIR}}}
\newcommand{\snr}{\ensuremath{{\sf SNR}}}

\newcommand{\BO}[1]{\ensuremath{O\paren{#1}}} 
\newcommand{\vecbold}[1]{\ensuremath{\boldsymbol{#1}}}

\renewcommand{\vec}[1]{\ensuremath{\boldsymbol{#1}}} 

\newcommand\myatop[2]{\genfrac{}{}{0pt}{}{#1}{#2}}
\IEEEoverridecommandlockouts
\begin{document}
\vspace{-0.0cm}
\title{A Tractable Framework for the Analysis of Dense Heterogeneous Cellular Networks}

\author{ Serkan Ak, Hazer Inaltekin,~\IEEEmembership{Member,~IEEE}, \\ H. Vincent Poor,~\IEEEmembership{Fellow,~IEEE} 
\thanks{S. Ak is with the Wireless Networking and Communications Group (WNCG), The University of Texas at Austin, Austin, TX 78701, USA (e-mail: serkanak@utexas.edu). H. Inaltekin is with the Department of Electrical and Electronic Engineering at the University of Melbourne, VIC 3010, Australia (e-mail: hazer.inaltekin@unimelb.edu.au). H. V. Poor is with the Department of Electrical Engineering, Princeton University, Princeton, NJ 08544, USA (e-mail: poor@princeton.edu).

This research was supported in part by the European Union Research Executive Agency Marie Curie FP7-Reintegration-Grants under Grant PCIG10-GA-2011-303713, in part by the Scientific and Technological Research Council of Turkey (TUBITAK) under Grant 115E162, and in part by the U.S. National Science Foundation under Grants CNS-1702808 and ECCS-1647198. This work was presented in part at the IEEE International Symposium on Information Theory (ISIT), Barcelona, Spain, July 2016 \cite{Ak1,Ak2}.
 } }
\maketitle
\vspace{-1.0cm}
\begin{abstract}
This paper investigates the downlink performance of dense $K$-tier heterogeneous cellular networks (HCNs) under general settings. First, Gaussian approximation bounds for the standardized aggregate wireless interference (AWI) in {\em dense} $K$-tier HCNs are obtained for when base stations (BSs) in each tier are distributed over the plane according to a spatial and general Poisson point process. The Kolmogorov-Smirnov (KS) distance is used to measure deviations of the distribution of the standardized AWI from the standard normal distribution.
An explicit and analytical expression bounding the KS distance between these two distributions is obtained as a function of a broad range of network parameters such as per-tier transmission power levels, per-tier BS intensity, BS locations, general fading statistics, and general bounded path-loss models. Bounds achieve
a good statistical match between the standardized AWI distribution and its normal approximation even for moderately dense HCNs. Second, various spatial performance metrics of interest such as outage capacity, ergodic capacity and area spectral efficiency in the downlink of $K$-tier HCNs for general signal propagation models are investigated by making use of the derived distribution approximation results. Considering two specific BS association policies, it is shown that the derived performance bounds track the actual performance metrics reasonably well for a wide range of BS intensities, with the gap among them becoming negligibly small for denser HCN deployments.  Finally, both analytical and numerical results on the area spectral efficiency reveal a non-linear growth trend with diminishing returns of HCN performance. Hence, the $\sir$ invariance property does not hold under bounded path-loss models, which is a critical finding from the HCN design perspective. In particular, it points out a critical BS density beyond which the HCN performance starts to decline due to excessive wireless interference.

\end{abstract}
\vspace{-0.0cm}

\section{Introduction}\label{Introduction}
\subsection{Background and Motivation}
Fifth generation (5G) wireless networks are conceived as highly heterogeneous consisting of multiple-tiers of network elements with much {\em denser} deployments, e.g., hundreds of transmitters per unit area and more advanced communication protocols to deal with excessive data demand from mobile users \cite{Andrews12, Ghosh12, Hwang13, Nguyen_Kountouris17, Ding_Lopez_Mao_17, Lopez15, Galinina15, Hanzo16, AlAmmouri17}. Modeling and analyzing performance of such multi-tier heterogeneous cellular networks (HCNs) using spatial point processes have recently gained increasing popularity \cite{Andrews11, Dhillon12, Jo12, SIE14a, Hossain13}.  Poisson cluster processes have also recently been used for the same purpose in \cite{Saha_Afshang_Dhillon17,Afshang_Dhillon17}.  In particular, it is shown in \cite{Andrews11} that using a Poisson point process (PPP) model even for macro cell BS locations provides us with an approximation as good as the one provided by the conventional grid based model \cite{Macdonald79} for the actual network performance.  The PPP based location model leads to a lower performance bound, whilst the grid model resulting in an upper performance bound, with almost the same deviations from the experimental data \cite{Andrews11}. The appropriateness of the random spatial models for network modeling and analysis is expected to increase even further with the more irregular topology of HCNs. 

Along with PPP based modeling for analytical tractability, there is an increasing tendency towards {\em network densification} in order to improve data rates in HCNs \cite{Hwang13, Nguyen_Kountouris17,Ding_Lopez_Mao_17, Lopez15, AlAmmouri17}.  From an engineering point of view, the significant capacity improvements are expected to emerge automatically in ultra-dense HCNs due to shorter connection distances arising from the deployment of hundreds (or, even thousands) of small cell base-stations (BSs) per square kilometer.  Contrary to this intuitive expectation, an important observation in some of these previous studies such as \cite{Nguyen_Kountouris17, Ding_Lopez_Mao_17, AlAmmouri17} is that there is a certain level of network densification to achieve the best data rate performance. It is also concluded in these papers that this critical level of network densification, depending on network modeling assumptions, is inherently related to a variety of networks parameters such as association policy, fading distribution, path-loss model, and even the antenna heights and elevation angles. 

Our results in this paper provide new insights into the performance gains that can be achieved through network densification in HCNs. Specifically, we prove that network interference resembles a Gaussian distributed random variable with a growing mean in dense HCNs under any given {\em bounded} path-loss model. Utilizing the discovered statistical structure of network interference across HCN performance analysis, it is also shown that network interference renders the advantages of network densification moot after some certain BS density level. Roughly speaking, this finding implies that it is the bounded nature of any path-loss model that will restrict the benefits of network densification strategies in next-generation wireless communication systems. Last but not least, the discovered Gaussian structure of network interference is also important from the network analysis and planning perspective as it facilitates the performance analysis of dense HCNs greatly by eliminating the need for complicated numerical integral computations for inverting complex-valued characteristic functions and/or Laplace transforms, as illustrated later in the paper through bounding various HCN capacity metrics.



More generally, a major design issue for PPP based HCN models, dense or not, is characterization and mitigation of aggregate wireless interference (AWI), e.g., see \cite{Seymour14} and \cite{Hossain14}. Although the statistical characterization of AWI is possible for the special case of Rayleigh fading and the classical unbounded path-loss model to compute various performance metrics such as outage probability in closed form \cite{Dhillon12, Jo12, SIE14a}, computation of the exact AWI distribution is a very challenging task that usually does not result in closed form expressions for general signal propagation models \cite{Andrews11, HG09}. In addition to network planning insights mentioned above, another important contribution of the current paper is to generalize those previously known results and techniques to study the HCN performance under more general and heterogeneous communication scenarios that take into account the general bounded path-loss models, general fading distributions and spatial distributions of BSs including non-homogeneous PPPs. This is achieved by leveraging our methodological analysis approximating the standardized AWI distribution as a normal distribution.   




\subsection{Main Contributions}

The main contributions of the current paper are listed below. 

\subsubsection{Development of a New Analytical Framework for the HCN Performance Analysis} 
This paper develops a new analytical framework for the HCN performance analysis.  The developed framework results in performance figures arbitrarily close to the actual network performance in the dense network limit.  Specifically, a thorough mathematical analysis of the AWI distribution under general network settings is often a prohibitively hard task. To overcome this difficulty, we introduce a simple and effective methodological approach for examining the statistical structure of AWI in the downlink of a {\em dense} $K$-tier HCN, wherein the network tiers are differentiated from each other in terms of transmission power levels, spatial BS distributions and RF signal propagation characteristics. 
The proposed approach provides us with a principled framework to analyze the downlink performance of spatially dense $K$-tier HCNs, where BS locations in each tier follow a PPP with possibly non-homogeneous spatial BS intensity.    
The signal power attenuation due to path-loss is modeled through a general bounded path-loss function decaying to zero with distance, which can vary from one tier to another. Fading and shadowing are also accounted for in the employed signal propagation model without assuming any specific distribution functions for these other random wireless channel dynamics. 
 
\subsubsection{Derivation of Gaussian Approximation Bounds for the Distribution of AWI in HCNs}
This paper derives tight Gaussian approximation bounds for the distribution of AWI for the downlink HCN communications.  More specifically, measuring the distance between the standardized downlink AWI and normal distributions by means of the Kolmogorov-Smirnov (KS) distance, we obtain an analytical expression for deviations between them, which paves the way for illuminating statistical behavior of the AWI and in turn for designing an efficient $K$-tier HCN. Briefly, the stated distance consists of two parts: (i) a scaling coefficient and (ii) a multiplicative positive function $c(x)$ with $x \in \R$ being the point at which we want to estimate the value of the standardized AWI distribution.  

The scaling coefficient has two important properties. First, it is related to the skewness characteristic of the individual interference terms constituting the standardized AWI, which mainly depends on various network parameters at each tier such as transmission powers, BS distribution and signal propagation characteristics. The second property is its monotonically decaying nature to zero with denser deployments of BSs per tier.  The function $c(x)$ is {\em uniformly bounded} by a small constant and approaches zero for large absolute values of $x$ at a rate $\abs{x}^{-3}$, which makes the derived bounds on the tails of the standardized downlink AWI distribution tight even for sparsely deployed HCNs.  
                  

\subsubsection{Derivation of the Bounds on Various HCN Performance Metrics Under General Settings}
Utilizing the derived Gaussian approximation results across HCN performance analysis, this paper produces analytical expressions for upper and lower bounds on performance metrics of interest in $K$-tier HCNs under general network settings and general PPPs. In particular, we focus on the downlink data rates in a {\em dense} $K$-tier HCN under two different BS association policies. We obtain tight performance bounds on the downlink outage capacity, ergodic capacity and area spectral efficiency ($\ASE$) in HCNs under both association policies for general signal propagation models. The derived bounds approximate the actual HCN performance metrics accurately for a wide range of BS intensities in each tier. The gap between bounds becomes negligibly small as the BSs are more densely deployed. 

\subsubsection{Network Design Insights}
This paper also produces new network design insights for $K$-tier HCNs. In particular, based on the derived analytical expressions, we show that the outage probability increases if the BSs in a $K$-tier HCN are more densely deployed. This finding implies that the celebrated signal-to-interference ratio ($\sir$) invariance property \cite{Andrews11, Dhillon12} does not hold anymore if a bounded path-loss model is used to characterize large-scale wireless propagation losses in the downlink of an HCN. Hence, there is a fundamental capacity limit beyond which the data rate performance of an HCN cannot be increased indefinitely by adding extra infrastructure. As a result, we must either mitigate interference more efficiently or find the optimum BS intensity per tier maximizing delivered data rates to mobile users in order to reap the benefits from network densification.    

Our simulations corroborate these findings by indicating that an increase in the network intensity induces degradations in outage capacity and ergodic capacity, each of which scales at least as fast as $\Theta \left( {\left\| \vecbold{\lambda} \right\|_2^{ -1}} \right)$, where $\vecbold{\lambda}  = {\left[ {{\lambda _1}, \ldots ,{\lambda _K}} \right]^{\top}}$ is the BS intensity vector. Further, the $\ASE$ performance of a $K$-tier HCN exhibits a non-linear growth trend with diminishing returns and a plateauing behavior after some BS density level.  From an HCN design perspective, this result further supports the necessity of setting BS intensities at each tier appropriately for the proper and cost-effective delivery of data services to mobile users.  Last but not least, the general path-loss model we consider in this paper covers the stretched exponential path-loss models and {\em bounded} multi-slope path-loss models  \cite{AlAmmouri17, Zhang15, Deng15}. Hence, our results are also appropriate for characterizing the performance of emerging millimeter wave (mmWave) communications in the dense network limit.  Overall, the proposed approach can be extended to other HCN settings and performance metrics, with the potential of providing new network design insights and shedding light on HCN performance beyond specific selections of the path-loss model and the fading distribution. To this end, some potential extensions of the proposed approach are further discussed in Section \ref{Section: Extentions}.


\subsection{Paper Organization and Notation}
The rest of the paper is organized as follows. In Section \ref{Related Work}, we present a survey of relevant previous work by comparing and contrasting existing results with those we obtain in the current paper.  In Section \ref{System_Model}, we introduce the system model. In Section \ref{Section: Gaussian_Approximation}, we present our methodological approach for approximating the AWI distribution in $K$-tier HCNs as a Gaussian distribution and illustrate numerical examples of these analytical findings. In Section \ref{Section: HCN Outage Performance}, we introduce BS association policies for performance analysis and derive our bounds tracking the performance metrics, especially outage characteristics of mobile users, under these association policies.  We illustrate simulation results on the derived performance bounds in Section \ref{Section: Simulation_Results}. We provide a discussion on the scope, generalizations and limitations of the derived results in Section \ref{Section: Extentions}. Finally, Section \ref{Conclusion} concludes the paper. 

We use boldface letters, upper-case letters and calligraphic letters to denote vector quantities, random variables and sets, respectively. $| \cdot |$ notation is used to measure the magnitudes of scalar quantities, whereas $\| \cdot \|_2$ notation is used to measure the Euclidean norms of vector quantities. $\I{\cdot}$ is used to denote the indicator function. Expected value and variance of a random variable $X$ are denoted by $\ES{X}$ and $\V{X}$, respectively.  
As is standard, when we write $f\left( t \right) = O\left( {g\left( t \right)} \right)$, $f\left( t \right) = \Omega \left( {g\left( t \right)} \right)$ and $f\left( t \right) = o\left( {g\left( t \right)} \right)$ as $t \to {t_0}$ for two positive functions $f\left( t \right)$ and $g\left( t \right)$, we mean $\lim {\sup _{t \to {t_0}}}\frac{{f\left( t \right)}}{{g\left( t \right)}} < \infty $, $\lim \inf{_{t \to {t_0}}}\frac{{f\left( t \right)}}{{g\left( t \right)}} > 0$ and ${\lim _{t \to {t_0}}}\frac{{f\left( t \right)}}{{g\left( t \right)}} = 0$, respectively. $f\left( t \right)$ is said to be $\Theta \left( {g\left( t \right)} \right)$ as ${t \to {t_0}}$ if $f\left( t \right) = O\left( {g\left( t \right)} \right)$ and $f\left( t \right) = \Omega \left( {g\left( t \right)} \right)$ as ${t \to {t_0}}$.

\section{Related Work}\label{Related Work}
The early work in the literature focusing on the design and analysis of wireless networks by means of stochastic geometry based models includes \cite{Sousa90, IH98, CH01, WA06, WPS09}. These papers considered traditional single-tier macro cell deployments and obtained various approximations on the distribution of AWI using characteristic functions \cite{Sousa90}, LePage series \cite{IH98}, Edgeworth expansion \cite{CH01}, geometrical considerations \cite{WA06} and skewed stable distributions \cite{WPS09}. In \cite{Sousa90}, a closed form expression was also obtained for the AWI distribution under the assumption of no fading and unbounded power-law decaying path-loss function when the path-loss exponent is $4$.  More recently, generalized shot-gun models are considered for one-, two- and three-dimensional wireless networks to derive semi-analytical expressions for the downlink coverage probability for arbitrary fading and general path-loss models in \cite{Madhu14}, optimum downlink coverage for Poisson cellular networks subject to transmit power, BS density and transmit power density constraints is derived in \cite{SIE14b}, and Berry-Esseen types of bounds were obtained in \cite{AY10, Inaltekin12, Inaltekin2-12}, but again by considering only single-tier wireless networks.  In \cite{Renzo_Lu15} and \cite{Afify15}, the authors used a Gaussian signal approximation technique to calculate bit error rate and symbol error rate as performance metrics for cellular networks. 

The current paper differs from the above previous work in several important aspects.  In particular, this paper extends the previous known results approximating the AWI distribution for macro cell deployments to more heterogeneous and complex wireless communication environments when compared to \cite{Sousa90, IH98, CH01, WA06, WPS09, Madhu14, SIE14b, AY10, Inaltekin12, Inaltekin2-12}. Functional dependencies among different tiers to approximate the AWI distribution in the downlink of a $K$-tier HCN are clearly identified. While the authors in \cite{WPS09} showed that the AWI distribution can be modeled as a skewed stable distribution with {\em unity} skew parameter under the unbounded path-loss model, our approximation technique uses general {\em bounded} path-loss models and demonstrates that the skewness of the standardized AWI distribution decreases to zero as the $K$-tier HCN under consideration gets denser.  We discover that the behavior of AWI changes from being heavy-tailed to an exponentially decaying light-tailed one for bounded path-loss models decaying to zero. 

This paper differs from \cite{Madhu14, SIE14b, AY10, Inaltekin12} by focusing on a multi-tier network scenario with distinct and general network parameters in each tier, which enables us to discover further insights into the behavior of AWI in dense HCNs via Lemmas \ref{Lemma: Convergence Rate} and \ref{Lemma: Effect of Fading} in Section \ref{Section: Gaussian_Approximation}.  The authors in \cite{Madhu14} focus only on {\em single-tier} networks, and their equivalence results are between one-, two- and three-dimensional Poisson wireless networks. As a result, the results presented in \cite{Madhu14} are not applicable to our case to first reduce the multi-tier network to a single-tier network, and then use the Gaussian approximation results presented in our previous work \cite{Inaltekin12, Inaltekin2-12}.  In comparison with the preliminary results presented in \cite{Inaltekin2-12}, the association policies (APs) studied for $K$-tier HCNs in Section \ref{Section: HCN Outage Performance} are much richer in terms of the network parameters that they include, which provides additional insights into various capacity metrics for dense $K$-tier HCNs.

The papers \cite{Renzo_Lu15} and \cite{Afify15} differ from the current paper by considering diversity at the bit and symbol level through bit-error-rate and/or symbol-error-rate calculations based on an equivalence-in-distribution (EiD) technique. The introduced EiD technique requires characteristic function (CF) calculations for the interference signal, while our Gaussian approximation approach does not require any such CF (and/or Laplace transform) calculations due the employed Berry-Esseen technique.  Further, we only consider the statistics of the AWI {\em power} in this paper, which is the mainstream approach taken in many previous studies, i.e., see \cite{HG09}. Amplitude statistics of the AWI considered in \cite{Renzo_Lu15} and \cite{Afify15} are strongly related to the type of signaling scheme employed, which is not within the scope of the current paper.

The related work also includes the papers that use stochastic geometry to model and analyze HCNs such as \cite{Renzo13, Sho15, Heath13, Mukherjee12,  SIE16, Madhu16, Blaszczyszyn15}. To start with, the papers \cite{Renzo13} and \cite{Sho15} calculate the statistics of the signal-to-interference-plus-noise ratio ($\sinr$) in the downlink of HCNs by utilizing moment generating functions (MGF). 
In \cite{Heath13}, the authors investigated a gamma distribution approximation for the distribution of AWI clogging a fixed-size cell with a guard zone and a dominant interferer. In \cite{Mukherjee12}, the author derived the downlink $\sinr$ distribution for $K$-tier HCNs by assuming the classical unbounded path-loss model, Rayleigh faded wireless links and the nearest base-station (BS) association rule. In \cite{SIE16}, they considered vector broadcast channels operating according to opportunistic beamforming in a $K$-tier HCN setting, and obtained tight approximations for beam outage probabilities and ergodic aggregate data rates for Rayleigh faded propagation environments with homogeneous PPP distributions for the locations of access points. 

In \cite{Madhu16}, the authors extended the previous results in \cite{Madhu14} for four different association policies in HCNs (i.e., max-$\sinr$, nearest-BS, maximum received instantaneous power and  maximum biased received power association models) for an unbounded path-loss model with specific functional form (having varying path-loss exponents from tier-to-tier) and with arbitrary fading when the BS locations in each tier are given by homogeneous PPPs. They obtained semi-analytical expressions for the downlink performance of HCNs involving complex-valued integrals. In \cite{Blaszczyszyn15}, the authors used the factorial moments for the $\sinr$ process to show that a generic $K$-tier HCN can be transformed into a stochastically equivalent single-tier network when the BSs in each tier are distributed over the plane according to homogeneous PPPs and the path-loss model is given by the unbounded inverse power-law function.   

The current paper differs from \cite{Renzo13} and\cite {Sho15} by utilizing general and tier-dependent bounded path-loss models as well as using general fading distributions. In comparison with wireless network performance results obtained for homogeneous PPPs, we show that incorporating the skewness of the standardized AWI via the Berry-Esseen theorem makes the Gaussian approximation analytically more useful and numerically more tractable to accurately approximate the AWI distribution even for non-homogeneous PPPs and general fading models. When compared with the results reported in \cite{Heath13, Mukherjee12, SIE16}, our network set-up is much richer, allowing non-homogeneous PPPs for BS locations and general signal propagation models including fading and shadowing. 

This paper differs from \cite{Madhu16} in two important aspects. Firstly, our network model is not restricted to homogeneous PPPs and the unbounded path-loss model.  Secondly, our Gaussian approximation technique does not necessitate an inversion of complex-valued characteristic functions.  More specifically, our model generalizes the results in \cite{Madhu16} to non-homogeneous PPPs with arbitrary mean-measure and to bounded path-loss models with arbitrary functional forms by only requiring computation of integrals with respect to the generalized fading distributions on the real line to obtain tight performance bounds on the downlink of $K$-tier HCNs. 
The approach presented in \cite{Blaszczyszyn15} for mapping a $K$-tier HCN to a stochastically equivalent single-tier planar wireless network is promising, but the key propagation invariance property disappears for general bounded path-loss models. Hence, such stochastic equivalence results are not directly applicable to the $K$-tier HCN setup with general bounded path-loss models and non-homogeneous PPPs studied in this paper. More generally, we observe that the $\sinr$ invariance property does not hold in our case, which reinforces the similar observations obtained for stretched exponential path-loss models and multi-slope path-loss models in \cite{Zhang15, AlAmmouri17}, respectively. 

\section{System Model}\label{System_Model}
In this section, we will introduce the details of the studied downlink model in a $K$-tier cellular topology, the details of the spatial processes determining BS locations, the signal propagation characteristics and the association policy under which the network performance of a $K$-tier HCN is determined.
\subsection{The Downlink Model in a $K$-Tier Cellular Topology} 
We consider an overlay $K$-tier HCN in which the BSs in all tiers are fully-loaded (i.e., no empty queues) and have access to the same communication resources both in time and frequency. The BSs in different tiers are differentiated mainly on the basis of their transmission powers, with $P_k > 0$ being the transmission power of a tier-$k$ BS for $k = 1, \ldots, K$. As is standard in stochastic geometric modeling, it is assumed that BSs are distributed over the plane according to a general PPP 
 with differing spatial density among the tiers. Further, the signal propagation characteristics (including both large-scale path-loss and small-scale fading) also vary from one tier to another. The details of BS location processes and signal propagation are elaborated below.  

We place a test user at an arbitrary point $\vecbold{x}^{(o)} = \paren{x^{(o)}_1, x^{(o)}_2} \in \R^2$ and consider signals coming from all BSs in all tiers as the {\em downlink} AWI experienced by this test user.  Since we focus on the downlink analysis, we assume that the uplink and downlink do not share any common communication resources.  Therefore, the uplink interference can be ignored for the analysis of downlink AWI. This setting is general enough to illuminate the effects of various network parameters such as transmission powers and BS intensity in each tier on the distribution of the AWI seen by the test user.

\subsection{BS Location Processes}
The BS locations in tier-$k$, $k=1, \ldots, K$, independently form a spatial planar PPP $\Phi_k$. $\Lambda_{k}$ represents the {\em mean measure} (alternatively called the intensity measure or spatial density) of the $k$th tier BSs.  We do not assume any specific functional form for $\Lambda_{k}$ and hence do not restrict our attention only to homogeneous PPPs. For each (Borel) subset $\mathcal{S}$ of $\R^2$, $\Lambda_{k}\paren{\mathcal{S}}$ gives us the average number of BSs lying in $\mathcal{S}$. We will assume that $\Lambda_{k}$ is {\em locally finite} i.e., $\Lambda_{k}\paren{\mathcal{S}} < \infty$ for all bounded subsets $\mathcal{S}$ of $\R^2$, and $\Lambda_{k}\paren{\R^2} = \infty$, i.e., there is an infinite population of tier-$k$ BSs scattered all around in $\R^2$. For the whole HCN, the aggregate BS location process, which is the superposition of all individual position processes, is denoted by ${\Phi } = \bigcup\nolimits_{k = 1}^K {\Phi_k}$. 

For mathematical convenience, we also express $\Phi_k$ as a discrete sum of Dirac measures as $ \Phi_k \paren{\mathcal{S}}  = \sum_{j \geq 1} \delta_{\vecbold{X}_j^{(k)}}\paren{\mathcal{S}}$, where $\delta_{\vecbold{X}_j^{(k)}}\paren{\mathcal{S}} = 1$ if $\vecbold{X}_j^{(k)} \in \mathcal{S} \subseteq \R^2$, and zero otherwise.  The level of AWI at $\vecbold{x}^{(o)}$ from tier-$k$ BSs depends critically on the distances between the points of $ \Phi_k $ and $\vecbold{x}^{(o)}$. It is well-known from the theory of Poisson processes that the transformed process $ \sum_{j \geq 1} \delta_{T\paren{\vecbold{X}_j^{(k)}}}$ is still Poisson (on the positive real line) with mean measure given by $\Lambda_{k} \circ T^{-1}$, where $T\paren{\vecbold{x}} = \left\| \vecbold{x} - \vecbold{x}^{(o)} \right\|_2 = \sqrt{\paren{x_1 - x^{(o)}_1}^2 +  \paren{x_2 - x^{(o)}_2}^2}$ and $T^{-1}\paren{\mathcal{S}} = \brparen{\vecbold{x} \in \R^2: T\paren{\vecbold{x}} \in \mathcal{S}}$ for all $\mathcal{S} \subseteq \R$ \cite{Kingman93}. We will assume that $\Lambda_{k} \circ T^{-1}$ has a density in the form $\Lambda_{k} \circ T^{-1}\paren{\mathcal{S}} = \lambda_k \int_{\mathcal{S}} \mu_k(t) dt$.  Here, $\lambda_k$ is a modeling parameter pertaining to the $k$th tier, which can be interpreted as the {\em BS intensity parameter}, that will enable us to control the average number of tier-$k$ BSs whose distances from $\vecbold{x}^{(o)}$ belong to $\mathcal{S}$ and interfere with the signal reception at the test user. 

In order to understand the model further, it is insightful to consider a homogeneous PPP with density $\lambda_k$ for tier-$k$ BSs. In this case, $\Lambda_k\paren{\mathcal{S}}$ is given by $\Lambda_k\paren{\mathcal{S}} = \lambda_k \cdot {\rm area}\paren{\mathcal{S}}$ and the specific functional form of $\mu_k(t)$ becomes equal to $\mu_k(t) = 2\pi t\I{t \geq 0}$, which can be easily obtained by transforming the points of the homogeneous PPP over $\R^2$ to $\R$ by using the Euclidean metric as a mapping between $\R^2$ and $\R$. This is the PPP model for BS locations frequently considered in most previous papers, e.g., see \cite{Andrews11,Dhillon12,Jo12}. In our model above, on the other hand, we allow the possibility of $\Lambda_k\paren{\mathcal{S}}$ to depend on the set $\mathcal{S} \subseteq \R^2$ through an arbitrary functional form, which is not necessarily translation invariant, as long as $\Lambda_k$ constitutes a locally finite measure over the subsets of $\R^2$. As a result, the functional form of $\mu_k(t)$ can assume any shape from the set of functions over $\R$, rather than being restricted to the specific form $\mu_k(t) = 2\pi t \I{t \geq 0}$. In this regard, $\lambda_k$ just functions as a tunable network parameter in order for us to obtain a dense HCN deployment. An alternative approach to obtain a dense HCN deployment is to vary the functional form of $\mu_k(t)$ itself. However, this approach does not result in a single parameter representation to plot the downlink AWI distributions and HCN performance metrics. The analytical results we derive in this paper hold for any functional form of $\mu_k(t)$ satisfying the above mild conditions on $\Lambda_k$.

\subsection{Signal Propagation Model}  
We model the large scale signal attenuation for tier-$k$, $k=1, \ldots, K$, by a {\em bounded} monotone non-increasing path-loss function $G_k: [0, \infty) \mapsto [0, \infty)$. $G_k$ asymptotically decays to zero at least as fast as $t^{-\alpha_k}$ for some path-loss exponent $\alpha_k > 2$.  To ensure the finiteness of AWI at the test user, we require the relationship ${\mu}_{k}(t) = \BO{t^{\alpha_k - 1 - \epsilon}}$ as $t \to \infty$ to hold for some $\epsilon > 0$.

The fading (power) coefficient for the wireless link between a BS located at point $\vecbold{X} \in \Phi$ and the test user is denoted by $H_{\vecbold{X}}$.\footnote{For simplicity, we only assign a {\em single} fading coefficient to each BS. In reality, it is expected that the channels between a BS and all potential receivers (intended or unintended) experience different (and possibly independent) fading processes. Our simplified notation does not cause any ambiguity here since we focus on the outage and rate performance of the test user at a given arbitrary position in $\mathbb{R}^{2}$ in the remainder of the paper.} The fading coefficients $\brparen{H_{\vecbold{X}}}_{\vecbold{X} \in \Phi}$ form a collection of independent random variables (also independent of ${\Phi}$), with those belonging to the same tier, say tier-$k$, having a common probability distribution with density $q_k(h), h \geq 0$.  The first, second and third order moments of fading coefficients are assumed to be finite, and are denoted by $m^{(k)}_{H}$, $m^{(k)}_{H^2}$ and $m^{(k)}_{H^3}$, respectively, for tier-$k$.  We note that this signal propagation model is general enough that $H_{\vecbold{X}}$'s could also be thought to incorporate {\em shadow fading} effects due to blocking of signals by large obstacles existing in the communication environment, although we do not model such random factors explicitly and separately in this paper. Further, the model can also be extended to the load-aware analysis of HCNs, multi-antenna communications and power-controlled BSs, as elaborated in Section \ref{Section: Extentions} of the paper.

\subsection{Association Policy, Interference Power and Performance Measures}
Association policy is a key mechanism that determines the outage and rate performance experienced by the test user as it regulates the useful signal power as well as the interference power at the test user. Hence, we first formally define it to facilitate the upcoming discussion. 

\begin{Def} \label{Def: Association Policy}
An association policy $\mathcal{A}: \Omega \times \Rp^\infty \times \Rp^K \times \Rp^K \mapsto \R^2$ is a mapping that takes a BS configuration $\varphi \in \Omega$ (i.e., a countable point measure), fading coefficients $\brparen{H_{\vecbold{x}}}_{\vecbold{x} \in \varphi}$, transmission power levels $\brparen{P_k}_{k=1}^K$ and biasing coefficients $\brparen{\beta_k}_{k=1}^K$ as an input and determines the BS location to which the test user is associated as an output.   
\end{Def}

For the HCN model explained above, the output of $\mathcal{A}$ is a random point $\vecbold{X}^\star  = \paren{X_1^\star, X_2^\star} \in \Phi$ since the BS locations and fading coefficients are random elements. Biasing coefficients are important design parameters to offload data from bigger cells to the smaller ones. Two other important random quantities related to $\vecbold{X}^\star$ are the tier index $A^\star$ to which $\vecbold{X}^\star$ belongs and the distance between $\vecbold{X}^\star$ and the test user $ \vecbold{x}^{(o)} = \paren{x_1^{(o)}, x_{2}^{(o)}} \in \R^2 $, which is denoted by $R^\star = \left\| \vecbold{X}^\star - \vecbold{x}^{(o)} \right\|_2 = \sqrt{\paren{X_1^\star - x_{1}^{(o)} }^2 + \paren{X_2^\star - x_{2}^{(o)} }^2}$. 
Using these definitions along with considering all the signal impairments due to fading and path-loss, the total interference power at the test user is written as
\begin{eqnarray} 
I_{\vecbold{\lambda}} = \sum_{\vecbold{X} \in \Phi \backslash \brparen{\vecbold{X}^\star}} P_{\vecbold{X}} H_{\vecbold{X}} G_{\vecbold{X}} \paren{T\paren{\vecbold{X}}}, \label{Eqn: Total Interference Power} 
\end{eqnarray}
 where $\vecbold{\lambda} = \sqparen{\lambda_1, \ldots, \lambda_K}^\top$, and it is understood that $P_{\vecbold{X}} = P_k$ and $G_{\vecbold{X}} = G_k$ if $\vecbold{X} \in \Phi_k$. 
This parametrization of AWI is chosen to emphasize the dependence of its distribution on the BS intensity parameter $\lambda_k$ of each tier.

$\sinr$ is the main performance determinant for the HCN model in question. Given an association policy $\mathcal{A}$, the $\sinr$ level experienced by the test user is equal to 
$$ \sinr_{\mathcal{A}} = \frac{P_{A^\star} H_{\vecbold{X}^\star} G_{A^\star}\paren{T\paren{\vecbold{X}^\star}} }{N_0 + \frac{1}{\PG} I_{\vecbold{\lambda}}},$$
where $N_0$ is the constant background noise power and $\PG \geq 1$ is a processing gain constant that signifies the interference reduction capability, if possible, of the test user.  We also let $\SNR_k = \frac{P_k}{N_0}$ to denote the signal-to-noise ratio ($\SNR$) for tier-$k$.  Next, we define the main performance metrics used to measure the HCN outage and rate performance.  
\begin{Def} \label{Def: Outage Metrics}
For a target bit rate $\tau$, $\tau$-outage probability is equal to 
$$\PRP{\tau\mbox{-outage}} = \PR{\log\paren{1 + \sinr_{\mathcal{A}}} < \tau}.$$ 
Similarly, for a target outage probability $\gamma$, the outage capacity  achieved by the test user under the association policy $\mathcal{A}$ is equal to
$$ C_{\rm o}\paren{\gamma} = \sup\brparen{\tau \geq 0: \PRP{\tau\mbox{-outage}} \leq \gamma},$$
which is the maximum data rate supported with outage probability not exceeding $\gamma$.\footnote{The focus is more on the outage probability, rather than the outage capacity, in most previous work such as \cite{Andrews11, Dhillon12, Jo12}. The outage probability and outage capacity are related performance metrics for HCN analysis, with an extra step required to calculate the outage capacity. The main reason for us to focus on the outage capacity in this paper is to have the same units with the ergodic capacity metric.}    
\end{Def}   

Unlike the outage capacity in which (the instantaneous) $\sinr_{\mathcal{A}}$ is assumed to be an unknown constant for the duration of channel coherence time, ergodic capacity is the average of (the instantaneous) capacity that can be achieved by averaging over a large number of coherence time intervals leading to Definition \ref{Def: Ergodic Metrics} below. 
\begin{Def} \label{Def: Ergodic Metrics}
The ergodic capacity achieved by the test user under the association policy $\mathcal{A}$ is equal to
$${C_{\rm erg}} = \EW[ \log\paren{1 + \sinr_{\mathcal{A}}} ].$$
\end{Def}

Another system performance metric we analyze in this paper is the area spectral efficiency ($\ASE$). In contrast to the outage capacity and ergodic capacity, the $\ASE$ metric captures the \textit{collective} network performance.  For the sake of simplicity, we will study the $\ASE$ for when BSs in each tier are homogeneously distributed.  Its formal definition is given below.
\begin{Def} \label{Def: Area Spectral Efficiency}
Consider a $K$-tier HCN in which the BSs in tier-$k$ are distributed according to a homogeneous PPP with intensity $\lambda_k$ for $k = 1, \ldots, K$.  Then, the $\ASE$, measured in terms of Nats/Sec/Hz/Area, achieved by the network is equal to
\begin{eqnarray*}
{\ASE}\paren{ {\vecbold{\lambda}},{\vecbold{\gamma}} } = \sum_{k = 1}^K {{\lambda _k}\left( {1 - {\gamma _k}} \right)} {C_{\rm o}}\left( {k,{\gamma _k}} \right),
\end{eqnarray*}
where $\vecbold{\lambda}  = {\left[ {{\lambda _1}, \ldots ,{\lambda _K}} \right]^{\top}}$ is the vector of BS intensities, $\vecbold{\gamma}  = {\left[ {{\gamma _1}, \ldots ,{\gamma _K}} \right]^{\top}}$ is the vector of target outage probabilities with $\gamma_k$, $k=1, \ldots, K$, being the target outage probability for tier-$k$ BSs, and $C_{\rm o} \paren{ {k,{\gamma _k}} }$ is the conditional outage capacity in tier-$k$, conditioned on the event that a mobile user is connected to a tier-$k$ BS. 
\end{Def} 

Since $\ASE$ measures the collective network performance, rather than the one observed at a point, it is assumed that all BSs in the network serve a mobile user. Hence, the above definition with conditional outage capacities makes sense.  In the next section, we  will first present our methodological approach establishing the Gaussian approximation bounds to measure the proximity of the AWI distribution to the normal distribution. These approximation results will be leveraged in Section \ref{Section: HCN Outage Performance} to obtain the outage and rate performance of HCNs.

\section{Gaussian Approximation for the AWI Distribution} \label{Section: Gaussian_Approximation}
In this section, we will establish the Gaussian approximation bounds under different spatial distribution assumptions for the standardized AWI distribution in the downlink of an HCN.\footnote{We do not assume any interference protection or cancellation to derive the results in this section.  In Section \ref{Section: HCN Outage Performance}, we show that the results in this section can be modified in a straightforward manner to incorporate the effects of association policies on the $K$-tier HCN performance.} Then, we will present numerical examples validating our theoretical work.    

\subsection{Analytical Results} \label{Gaussian_Approximation_Theoretical_Work}
In this part, we present our theorems providing explicit upper and lower bounds on the AWI distribution, which the test user experiences at a specific location in a $K$-tier HCN. These bounds will clearly show the functional dependence between the downlink AWI distribution and a broad range of network parameters such as transmission power levels, BS distribution over the plane and signal propagation characteristics in each tier. We will also specialize these approximation results to the commonly used homogeneous PPPs at the end of this part. 

\begin{thm} \label{Theorem_1} 
For all $x \in \mathbb{R}$,
\begin{eqnarray}
\abs{\PR{\frac{I_{\vecbold{\lambda}} - \ES{I_{\vecbold{\lambda}}}}{\sqrt{\VS{I_{\vecbold{\lambda}}}}} \leq x} - \Psi(x)} \leq \Xi \cdot c(x),
\end{eqnarray}
where $\Xi = \sum_{k=1}^K \frac{\lambda_k P_k^3 m_{H^3}^{(k)} \int_0^\infty G_k^3(t) \mu_k(t) dt}{\paren{\sum_{k=1}^K \lambda_k P_k^2 m_{H^2}^{(k)} \int_0^\infty G_k^2(t) \mu_k(t) dt}^\frac32}$, $c(x) = \min\paren{0.4785, \frac{31.935}{1 + \abs{x}^3}}$ and $\Psi(x) = \frac{1}{\sqrt{2\pi}}\int_{-\infty}^x \e{- \frac{t^2}{2}} dt$, which is the standard normal cumulative distribution function (CDF). 
\end{thm}
\begin{IEEEproof}
We only give a sketch of the proof. The complete details can be found in \cite{Ak17}. There are several critical steps involved in the proof of this theorem. First, we show that the Laplace transform ${{\cal L}_{{I_{\vec{\lambda}} }}}\left( s \right) = \ES{ {\exp \left( { - s{I_\lambda }} \right)} }$ for $I_{\vec{\lambda}}$ exists for all $s \ge 0$ under our modeling assumptions in Section \ref{System_Model}. Second, we introduce a sequence of auxiliary i.i.d. random variables $U_{1,n}^{\left( k \right)}, \ldots ,U_{\left\lceil {{\Lambda _{n,k}}} \right\rceil ,n}^{\left( k \right)}$ for each $n \in \mathbb{N}$ and for each tier $k \in \left\{ {1, \ldots ,K} \right\}$, with a common PDF ${f_k} \left( t \right) = \frac{{{\lambda _k}{\mu _k}\left( t \right)}}{{{\Lambda _{n,k}}}}{1_{\left\{ {0 \le t \le n} \right\}}}$ for tier-$k$, where ${\Lambda _{n,k}} = {\lambda _k}\int_0^n {{\mu _k}\left( t \right)dt}$ and $\left\lceil . \right\rceil$ is the smallest integer greater than or equal to its argument. We define ${I_n} = \sum\nolimits_{k = 1}^K {I_n^{\left( k \right)}} $, where $ I_n^{\left( k \right)} = {P_k}\sum\nolimits_{i = 1}^{\left\lceil {{\Lambda _{n,k}}} \right\rceil } {H_i^{\left( k \right)}} {G_k}\left( {U_{i,n}^{\left( k \right)}} \right)$ and $\left\{ {H_i^{\left( k \right)}} \right\}_{i = 1}^\infty $ is an i.i.d. collection of random variables with the common probability density function $q_k\left(h\right)$ for $k = 1, \ldots ,K$. We prove
that $I_n$ converges in distribution to $I_{\vec{\lambda}}$ by showing that ${{\cal L}_{{I_n }}}\left( s \right)$ converges to ${{\cal L}_{{I_{\vec{\lambda}} }}}\left( s \right)$ pointwise as $n$ tends to infinity. Third, we obtain a bound on the deviations of the distribution of the standardized version of $I_n$, i.e., $\frac{{{I_n} - \ES{ I_n }} }{{\sqrt { \V{ I_n } } }}$, from a standard normal distribution by using the form of the Berry-Esseen theorem given in \cite{Chen05}. Finally, by combining the results proven in the second and third steps, we conclude the proof.    
\end{IEEEproof}

Measuring the KS distance, Theorem \ref{Theorem_1} provides us with an explicit expression for the deviations between the standardized AWI and normal distributions. Several important remarks about this result are in order. First, the standardized AWI can take negative values due to the centering operation, which makes the deviations from the normal distribution bounded for negative values of $x$, as desired in Theorem \ref{Theorem_1}.  Second, since a bounded path-loss model is used in each tier, the decay rate of the tails of the AWI distribution also depends on the fading distribution parameters in each tier as indicated by our bound in Theorem \ref{Theorem_1}.  A similar phenomenon was also observed in \cite{Inaltekin09, HG09}.  
We refer interested readers to \cite{HG09, Inaltekin09} for a thorough comparison between bounded and unbound path-loss models.  Third, the scaling coefficient $\Xi$ appearing in Theorem \ref{Theorem_1} is linked to the main network parameters such as transmission power levels, distribution of BSs over the plane and signal propagation characteristics. Starting with the BS intensity parameters $\lambda_k$, $k = 1, \ldots, K$, we observe that the rate of growth of the expression appearing in the denominator of $\Xi$ is half an order larger than that of the expression appearing in the numerator of $\Xi$ as a function of $\lambda_k$. This observation implies that the derived Gaussian approximation becomes tighter for denser deployments of HCNs.  A formal statement of this result is given in Lemma \ref{Lemma: Convergence Rate}.  

Fourth, the same functional form for $c(x)$ was also obtained in papers \cite{Inaltekin12, Inaltekin2-12}.  The reason is the fact that the same form of the Berry-Esseen theorem from \cite{Chen05} is applicable to both single-tier and multi-tier networks, but not the stochastic equivalence between a single-tier network and multi-tier HCNs \cite{Madhu16, Blaszczyszyn15}.  Fifth, an approach alternative to our Gaussian approximation method in Theorem \ref{Theorem_1} would be to use characteristic functions and/or Laplace transforms in an inversion formula to derive performance metrics for $K$-tier HCNs. However, this is not tractable for the HCN setup with non-homogeneous PPPs, general path-loss models and general fading distributions considered in this paper.  For example, an important feature of our Gaussian approximation result in Theorem \ref{Theorem_1} is that it only depends on the fading distributions through their second and third moments.  On the other hand, fading distributions appear in a more convoluted manner in the method of characteristic functions and/or Laplace transforms that necessitates the computation of integrals with respect to both the mean measure of the underlying BS processes and the per-tier fading distributions.  Finally, the skewness of the standardized AWI is related to its third moment, whose absolute value is upper bounded by $\Xi$ in Theorem \ref{Theorem_1}, and hence a decreasing function of each BS intensity.        


\begin{lem} \label{Lemma: Convergence Rate}
The scaling coefficient $\Xi$ appearing in the Gaussian approximation result in Theorem \ref{Theorem_1} is bounded above by $\Xi \leq \frac{\delta}{\sqrt{\| \vecbold{\lambda} \|_2}}$ for some finite positive constant $\delta$.   
\end{lem}       
\begin{IEEEproof}
Let $a_k = P_k^3 m_{H^3}^{(k)} \int_0^\infty G_k^3(t) \mu_k(t) dt$ and $b_k = P_k^2 m_{H^2}^{(k)} \int_0^\infty G_k^2(t) \mu_k(t) dt$. Then, 
\begin{eqnarray}
\Xi &=& \frac{\sum_{k=1}^K a_k \lambda_k}{\paren{\sum_{k=1}^K \lambda_k b_k}^\frac32} \leq \frac{\| \vecbold{\lambda}\|_2 \| \vecbold{a} \|_2}{\paren{\sum_{k=1}^K \lambda_k b_k}^\frac32} \nonumber
\end{eqnarray}
due to the Cauchy-Schwarz inequality. Further, we can lower-bound the sum in the denominator above as
\begin{eqnarray}
\paren{\sum_{k=1}^K \lambda_k b_k}^\frac32 \geq \paren{\min_{1 \leq k \leq K} b_k \sum_{k=1}^K \abs{\lambda_k}}^\frac32 \geq \epsilon \paren{\| \vecbold{\lambda} \|_2}^\frac32, \nonumber
\end{eqnarray}
where the last inequality follows from the equivalence of all the norms in finite dimensional vector spaces. Combining these two inequalities, we conclude the proof. 
\end{IEEEproof}

In the case of ultra-dense HCNs, the asymptotic behavior of the scaling coefficient $\Xi$ (and the related skewness behavior) is formally described by Lemma \ref{Lemma: Convergence Rate}, i.e., $\Xi = \BO{\frac{1}{\sqrt {\| \vecbold{\lambda} \|_2} }}$ as $\| \vecbold{\lambda} \|_2$ goes to infinity. 
Further, the effect of an association policy on $\delta$ in Lemma \ref{Lemma: Convergence Rate} can be seen more clearly by considering specific association policies (i.e., see Lemma \ref{Lemma: Gauss Approximation for Generic Policy} or Lemma \ref{Lemma: Gauss Approximation for BARSS Policy}).  In particular, the association policies considered in this paper modify the lower limits of the integrals in the numerator and denominator of $\Xi$. The numerical value of the ratio of these integrals are usually small numbers a little greater than one, i.e., see \cite{Inaltekin12} for one specific example. Moreover, the existence of the multiplying function $c(x)$ makes our bounds further sharpened, and hence the effect of a specific association policies on our Gaussian approximation stays limited.  

Following a similar approach above, we can also see that changing transmission powers is not as effective as changing BS intensity parameters to improve the Gaussian approximation bound in Theorem \ref{Theorem_1}. This is expected since the power levels are assumed to be deterministic (i.e., no power control is exercised) and therefore they do not really add to the randomness coming from the underlying spatial BS distribution over the plane and the path-loss plus fading characteristics modulating transmitted signals.  Another important observation we have in regards to the combined effect of the selection of transmission powers per tier and the moments of fading processes in each tier on the Gaussian approximation result in Theorem \ref{Theorem_1} is that our approximation bounds benefit from the fading distributions with restricted dynamic ranges and the alignment of received AWI powers due to fading and path-loss components.  This observation is made rigorous through the following lemma.    
\begin{lem} \label{Lemma: Effect of Fading}
Let $a_k = \lambda_k \int_0^\infty G_k^3(t) \mu_k(t) dt$, $b_k = \lambda_k \int_0^\infty G_k^2(t) \mu_k(t) dt$ and $c_k = P_k^2 m_{H^2}^{(k)}$.  Then, the scaling coefficient $\Xi$ appearing in the Gaussian approximation result in Theorem \ref{Theorem_1} is bounded below by 
$$ \Xi \geq \paren{\frac{1}{\|\vecbold{c}\|_2 \| \vecbold{b} \|_2}}^\frac32 \sum_{k=1}^K a_k c_k^\frac32,$$
with equality achieved if fading processes in all tiers are deterministic and the vectors $\vecbold{b} = \sqparen{b_1, \ldots, b_K}^\top$ and $\vecbold{c} = \sqparen{c_1, \ldots, c_K}^\top$ are parallel.  
\end{lem}
\begin{IEEEproof}
Using $a_k, b_k$ and $c_k$ introduced above, we can write a lower bound for $\Xi$ as
\begin{eqnarray*}
\Xi = \frac{\sum_{k=1}^K a_k P_k^3 m_{H^3}^{(k)}}{\paren{\sum_{k=1}^K b_k c_k}^\frac32} &=&  \frac{\sum_{k=1}^K a_k P_k^3 m_{H^3}^{(k)}}{\paren{\| \vecbold{c} \|_2}^\frac32 \paren{\sum_{k=1}^K b_k \frac{c_k}{\| \vecbold{c}\|_2}}^\frac32} \\
&\geq& \paren{\frac{1}{\| \vecbold{c} \|_2 \| \vecbold{b} \|_2}}^\frac32 \sum_{k=1}^K a_k P_k^3 m_{H^3}^{(k)}.
\end{eqnarray*}
Using Jensen's inequality, we also have $m_{H^3}^{(k)} \geq \paren{m_{H^2}^{(k)}}^\frac32$. Using this lower bound on $m_{H^3}^{(k)}$ in the above expression, we finally have $ \Xi \geq \paren{\frac{1}{\| \vecbold{c} \|_2 \| \vecbold{b} \|_2}}^\frac32 \sum_{k=1}^K a_k c_k^\frac32$. 
\end{IEEEproof}

In addition to the above fundamental properties of the scaling coefficient $\Xi$, it is also worthwhile to mention that the Gaussian approximation bound derived in Theorem \ref{Theorem_1} is a combination of two different types of Berry-Esseen bounds embedded in the function $c(x)$. One of these bounds is a {\em uniform} bound that helps us to estimate the standardized AWI distribution uniformly as
$$ \abs{\PR{\frac{I_{\vecbold{\lambda}} - \ES{I_{\vecbold{\lambda}}}}{\sqrt{\VS{I_{\vecbold{\lambda}}}}} \leq x} - \Psi(x)} \leq \Xi \cdot 0.4785 $$
for all $x \in \R$.  On the other hand, the other one is a {\em non-uniform} bound that helps us to estimate the {\em tails} of the standardized AWI distribution as
$$ \abs{\PR{\frac{I_{\vecbold{\lambda}} - \ES{I_{\vecbold{\lambda}}}}{\sqrt{\VS{I_{\vecbold{\lambda}}}}} \leq x} - \Psi(x)} \leq \Xi \cdot \frac{31.935}{1+\abs{x}^3}$$ 
and decays to zero as a third order inverse power law. 

Up to now, we considered general PPPs for the distribution of BSs in each tier. One simplifying assumption in the literature is to assume that PPPs determining the locations of BSs are homogeneous. In this case, $\mu_k(t)$ for all tiers is given by $\mu_k(t) = 2\pi t \I{t \geq 0}$, where $\I{\cdot}$ is the indicator function. Using this expression for $\mu_k(t)$ in Theorem \ref{Theorem_1}, we obtain the following approximation result for the distribution of AWI when all BSs are homogeneously distributed over the plane according to a PPP with differing BS intensity parameters $\lambda_k$ from tier to tier.   
 
\begin{thm} \label{Theorem_2} 
Assume that $\Phi_k$ is a homogeneous PPP with a mean measure given $\Lambda_{k}\paren{\mathcal{S}}  = \lambda_k \cdot \mbox{area}\paren{\mathcal{S}}$. Then, for all $x \in \R$,
\begin{eqnarray}
\abs{\PR{\frac{I_{\vecbold{\lambda}} - \ES{I_{\vecbold{\lambda}}}}{\sqrt{\VS{I_{\vecbold{\lambda}}}} } \le x} - \Psi(x)} \leq \Xi \cdot c(x),
\end{eqnarray}
where $\Xi = \frac{1}{\sqrt{2\pi}} \sum_{k=1}^K \frac{\lambda_k P_k^3 m_{H^3}^{(k)} \int_0^\infty G_k^3(t) t dt}{\paren{\sum_{k=1}^K \lambda_k P_k^2 m_{H^2}^{(k)} \int_0^\infty G_k^2(t) t dt}^\frac32}$, $c(x) = \min\paren{0.4785, \frac{31.935}{1+\abs{x}^3}}$ and $\Psi(x) = \frac{1}{\sqrt{2\pi}}\int_{-\infty}^x \e{- \frac{t^2}{2}} dt$, which is the standard normal CDF.    
\end{thm}
\begin{IEEEproof}
The proof follows from Theorem \ref{Theorem_1} by replacing $\mu_k(t)$ with $2\pi t \I{t \geq 0}$.
\end{IEEEproof}

When all network parameters are assumed to be the same, i.e., the same transmission power levels, fading distributions and BS distributions for all tiers, the HCN in question collapses to a single tier network. In this case, the Gaussian approximation result is given below.  
\begin{cor}\label{Corollary_1}
Assume $P_k = P$, $\mu_k(t) = 2\pi t \I{t \geq 0}$, ${G_k}\left( t \right) = G\left( t \right)$, $\lambda_k = \lambda$, $m_{H^2}^{(k)} = m_{H^2}$ and $m_{H^3}^{(k)} = m_{H^3}$ for all $k=1, \ldots, K$. Then, for all $x \in \R$, we have
$$ \abs{\PR{\frac{I_{\vecbold{\lambda}} - \ES{I_{\vecbold{\lambda}}}}{\sqrt{\VS{I_{\vecbold{\lambda}}}}} \le x} - \Psi(x)} \leq \Xi \cdot c(x), $$
where $\Xi = \frac{1}{\sqrt{2 \pi}} \frac{1}{\sqrt{K \lambda}} \frac{m_{H^3}}{\paren{m_{H^2}}^\frac32} \frac{\int_0^\infty G^3(t) t dt}{\paren{\int_0^\infty G^2(t) t dt}^\frac32}$, and $c(x)$ and $\Psi(x)$ are as given in Theorem \ref{Theorem_1}. 
\end{cor}

We note that this is the same result obtained in \cite{Inaltekin12} as a special case of the network model studied in this paper. 

\subsection{Numerical Verification of the Gaussian Approximation Results}\label{Verification of Gaussian Approximation Through Simulations}

\begin{figure*}[!t]
\begin{minipage}[b]{0.45\linewidth} 
\centering
\hspace{0.2cm}
\includegraphics[width=3.35in]{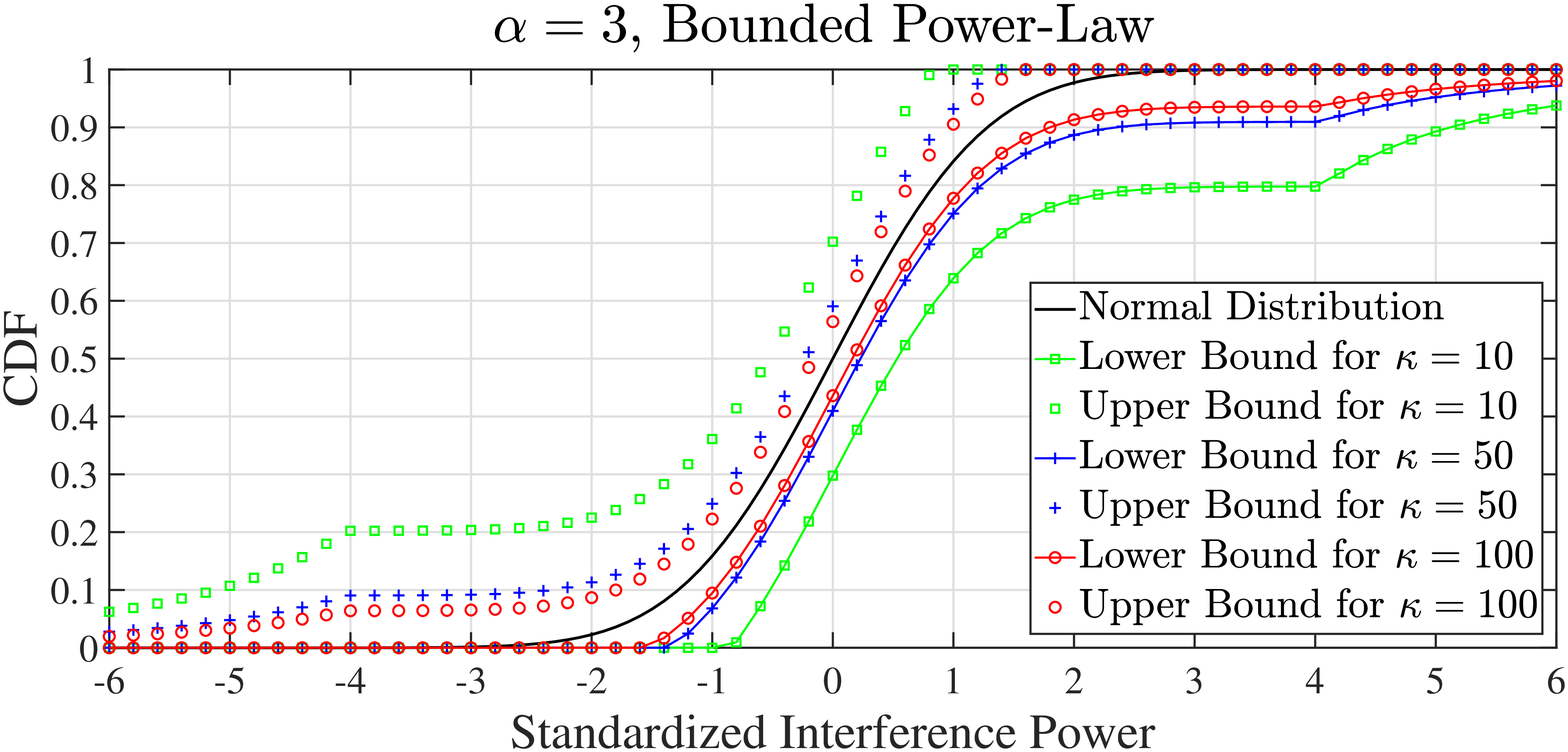}
\end{minipage}
\hspace{0.6cm} 
\begin{minipage}[b]{0.45\linewidth}
\centering
\includegraphics[width=3.35in]{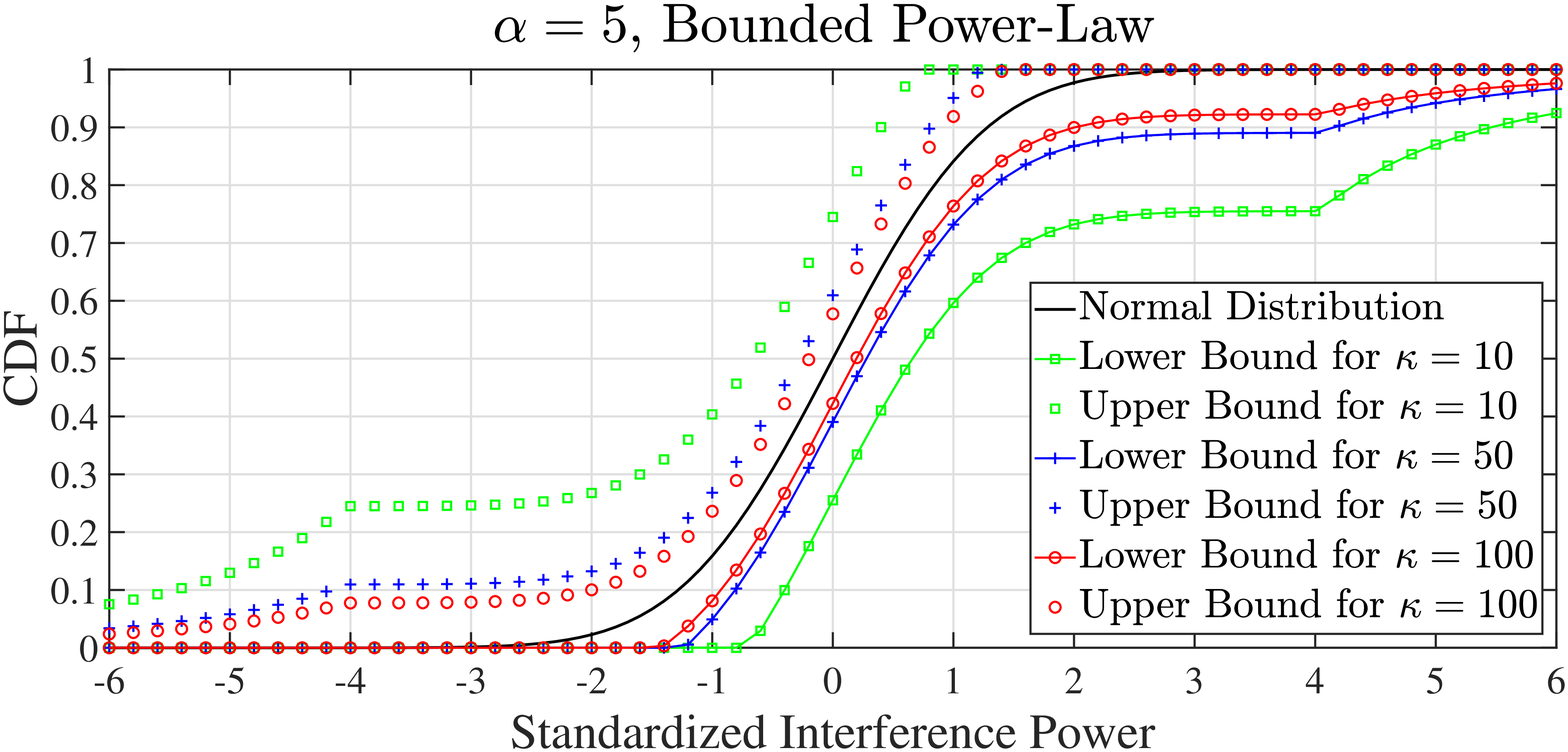}
\end{minipage}
\\
\begin{minipage}[b]{0.45\linewidth} 
\centering
\includegraphics[width=3.35in]{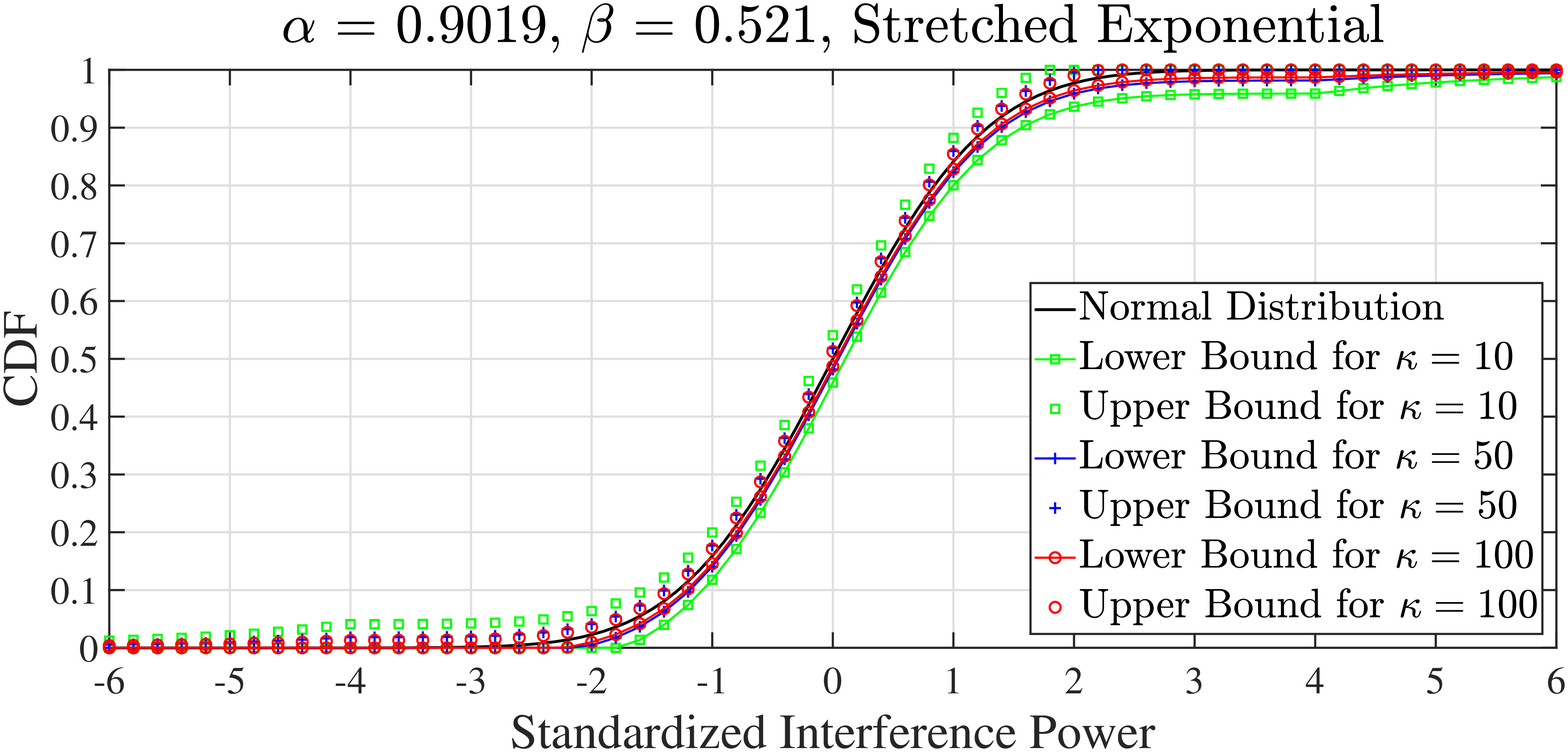}
\end{minipage}
\hspace{0.6cm} 
\begin{minipage}[b]{0.5\linewidth}
\centering
\includegraphics[width=3.35in]{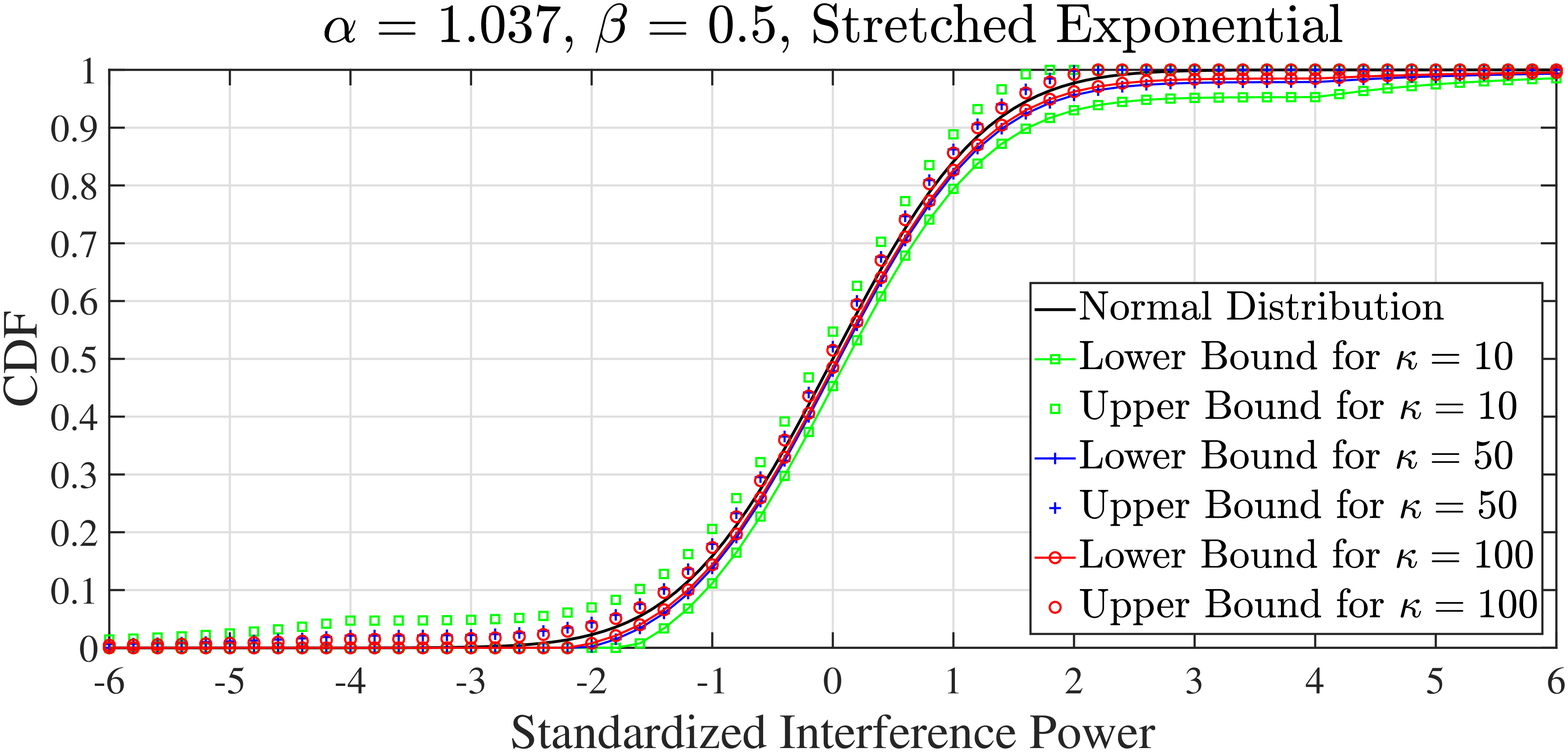}
\end{minipage}
\begin{minipage}[b]{0.45\linewidth} 
\centering
\includegraphics[width=3.35in]{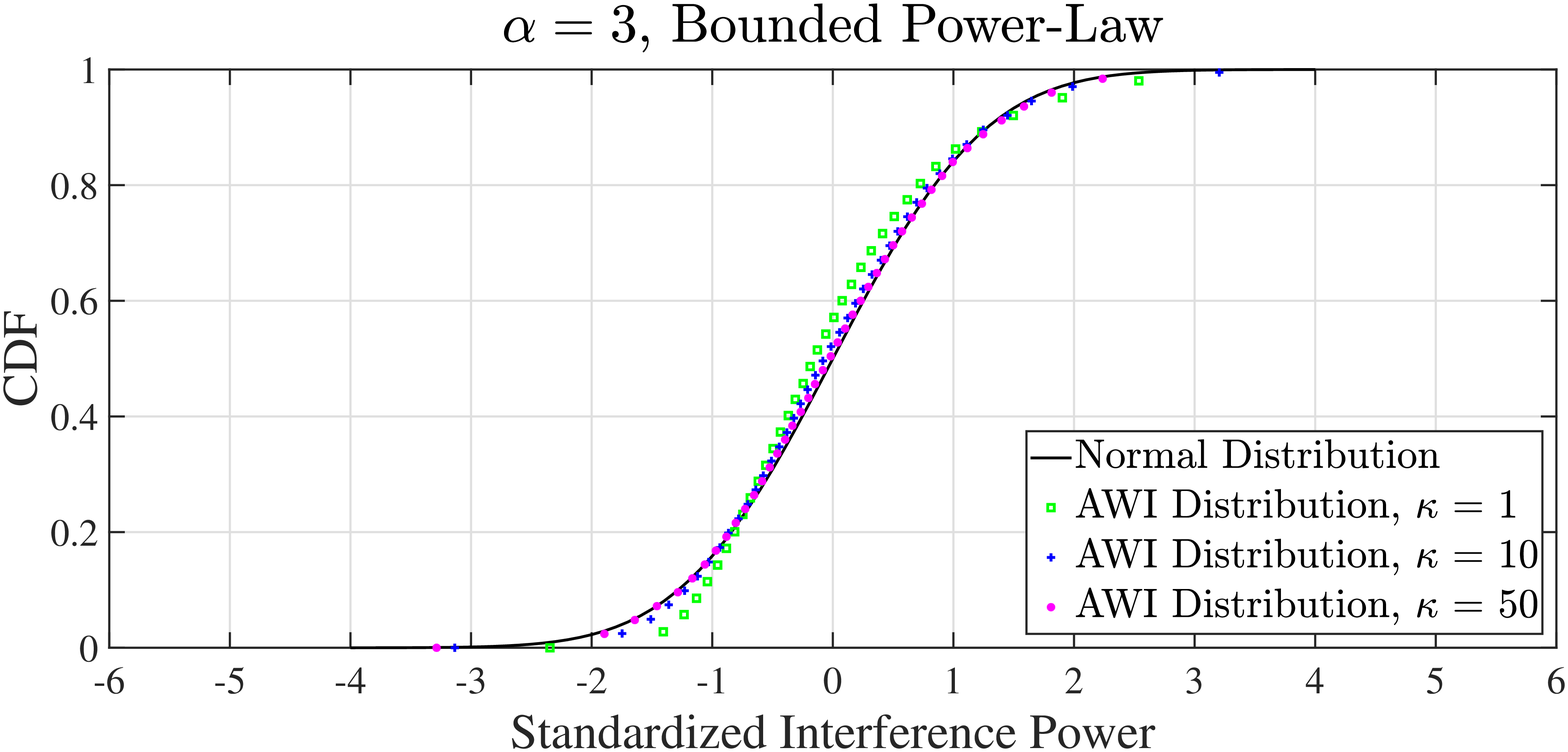}
\end{minipage}
\hspace{0.6cm} 
\begin{minipage}[b]{0.45\linewidth}
\centering
\includegraphics[width=3.35in]{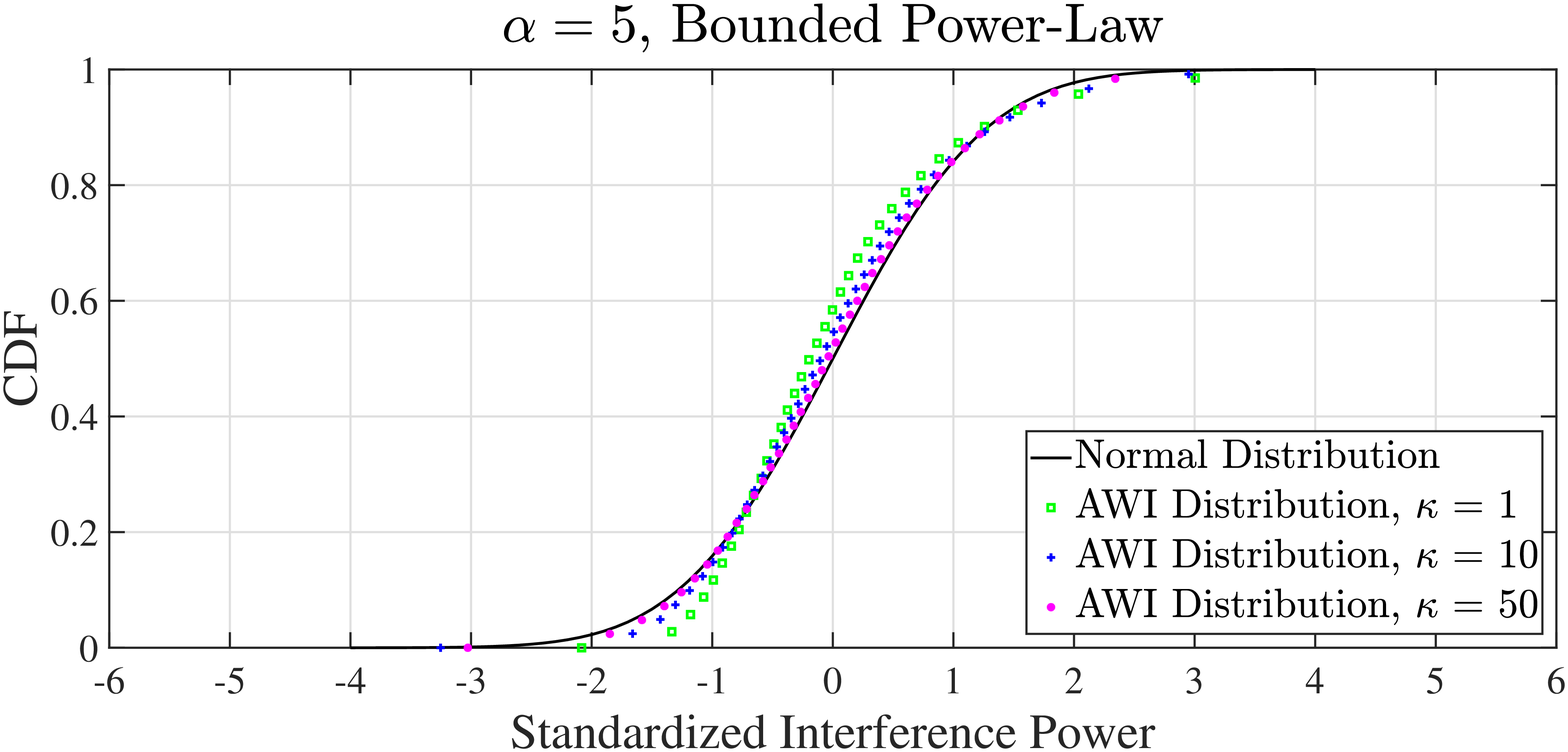}
\end{minipage}
\caption{Gaussian approximation bounds for the standardized AWI CDFs (top and middle figures).  Comparison of the simulated standardized AWI CDFs with the standard normal CDF (bottom figures). Parameter selection for the stretched exponential distribution is taken from \cite{AlAmmouri17}.  Rayleigh fading with unit mean power is assumed.} 
\label{fig1}
\end{figure*}



In this part, we will illustrate the analytical Gaussian approximation results derived for the standardized AWI distribution in Section \ref{Gaussian_Approximation_Theoretical_Work} for a specific three-tier HCN scenario. To this end, we will use a bounded power-law path-loss model $G\paren{t} = \frac{1}{{1 + {t^\alpha }}}$ and the recently proposed stretched exponential path-loss model $G\paren{t} = \exp\paren{-\alpha t^\beta}$ for dense cellular networks \cite{AlAmmouri17}.\footnote{With a slight abuse of notation, we adopted the same notation from \cite{AlAmmouri17} in order the represent the stretched exponential path-loss model. Hence, $\beta$ parameter in this path-loss model should not confused with the biasing coefficients introduced in Section \ref{System_Model}.} The path-loss models are assumed to be the same for all tiers with various values of the parameters $\alpha, \beta > 0$, and hence we do not index the path-loss models with the subscript $k$ below. Similar conclusions continue to hold for other path-loss models.  

The BSs in each tier are distributed over the plane according to a PPP, with BS intensity parameters given by $\lambda _1 = 0.1\kappa$, $\lambda _2 = \kappa$ and $\lambda _3 = 5\kappa$. Here, $\kappa$ is our unit-less control parameter to control the average number of BSs interfering with the signal reception at the test user.  The test user is assumed to be located at the origin without loss of any generality. 
The random fading coefficients in all tiers are assumed to be independent and identically distributed random variables, drawn from a Rayleigh distribution with unit mean power gain. Our results are qualitatively the same for other fading distributions such as Nakagami and Rician fading distributions.  The transmission power levels are set as $P_1 = 4 P_2 = 16 P_3$, where $P_2$ is assumed to be unity. 

\begin{figure*}[!t]
\begin{minipage}[b]{0.45\linewidth} 
\centering
\hspace{0.2cm}
\includegraphics[width=3.35in]{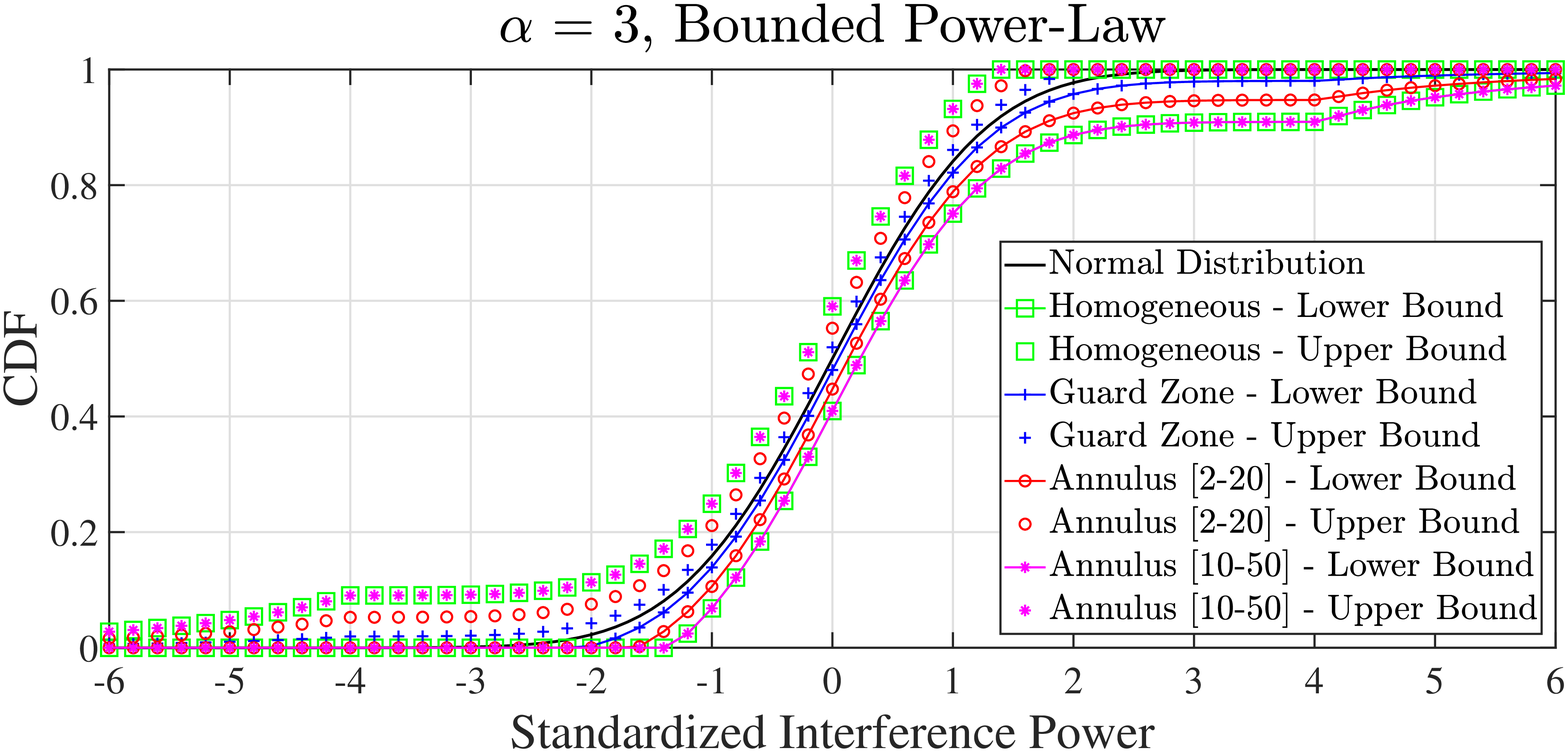}
\end{minipage}
\hspace{0.6cm} 
\begin{minipage}[b]{0.45\linewidth}
\centering
\includegraphics[width=3.35in]{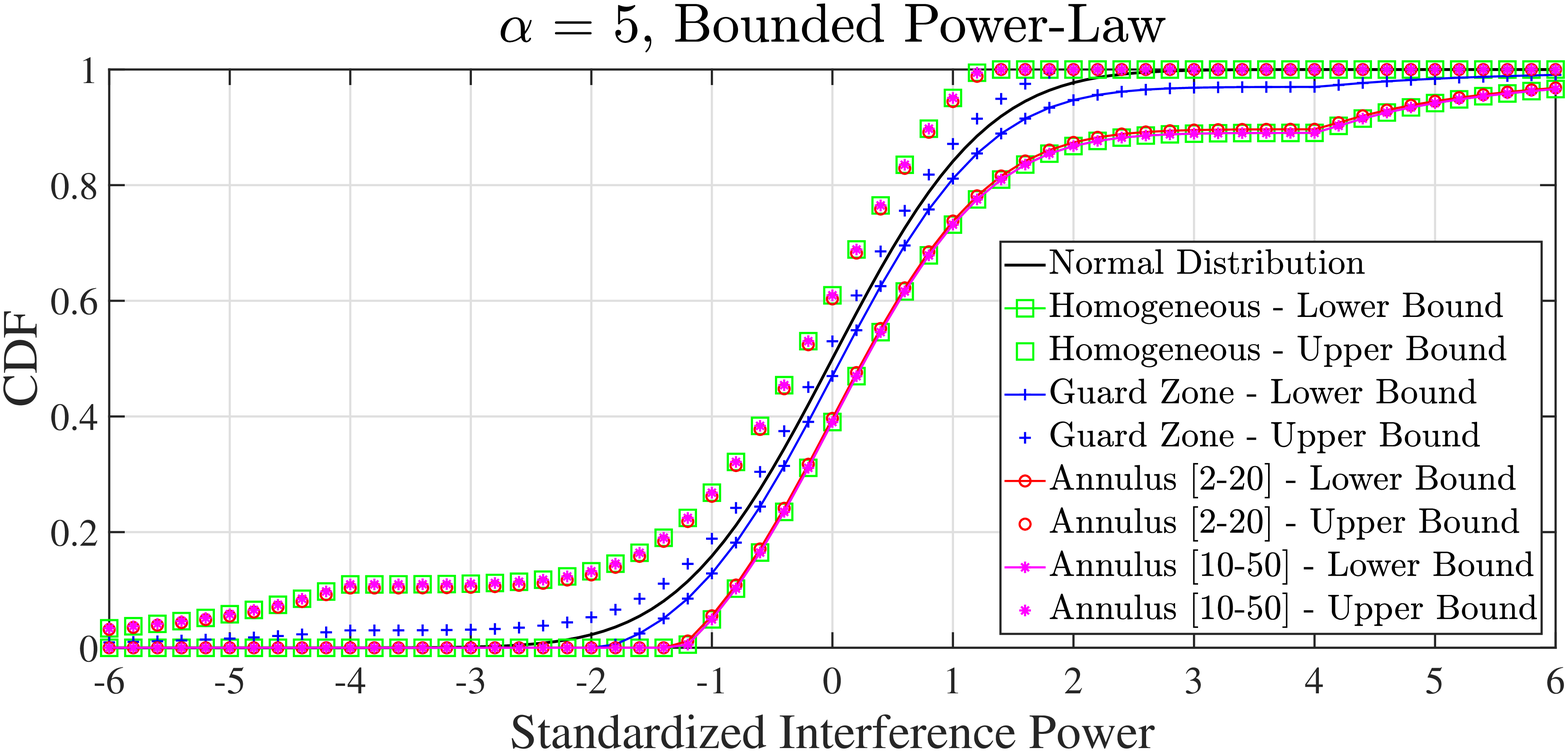}
\end{minipage}
\\
\begin{minipage}[b]{0.45\linewidth} 
\centering
\includegraphics[width=3.35in]{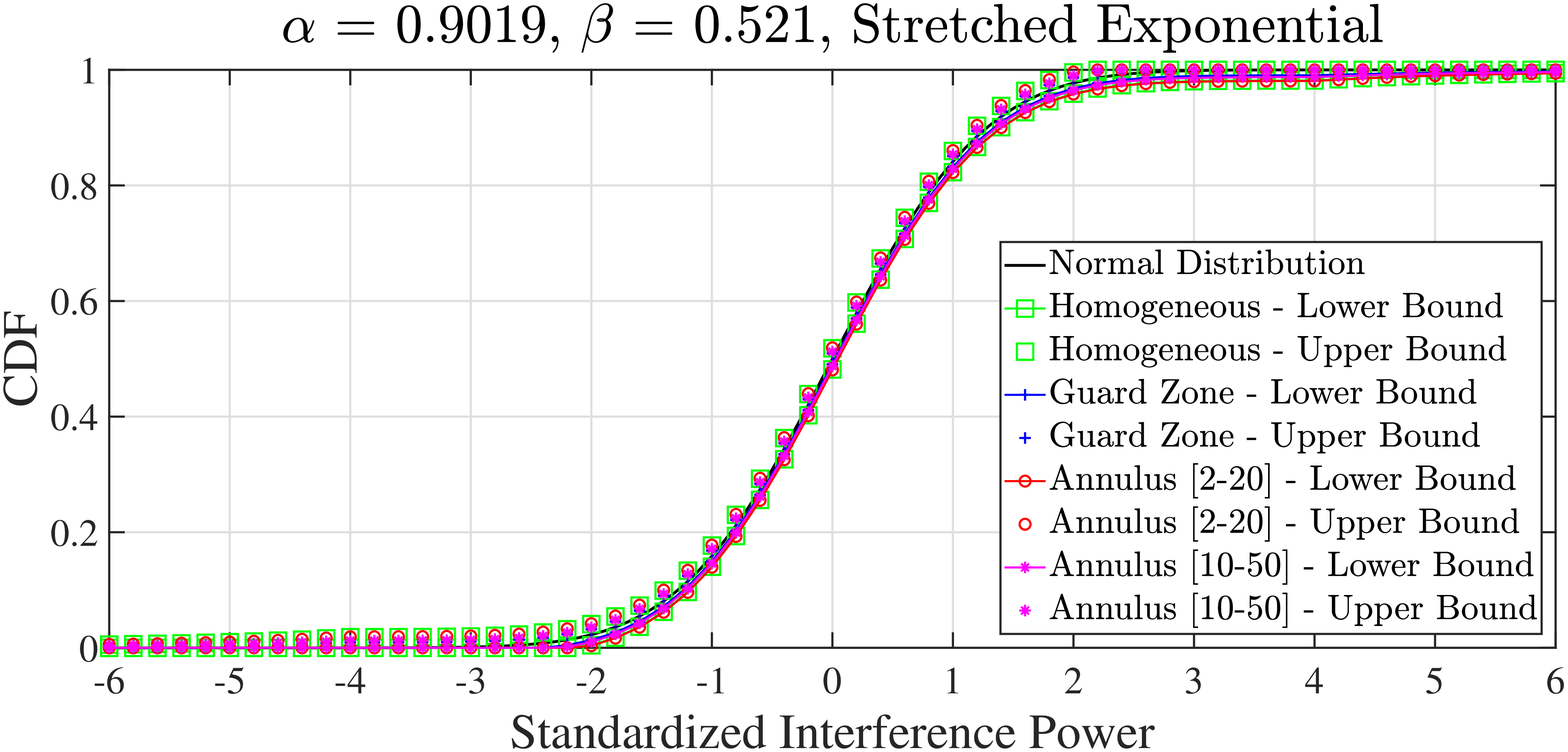}
\end{minipage}
\hspace{0.6cm} 
\begin{minipage}[b]{0.5\linewidth}
\centering
\includegraphics[width=3.35in]{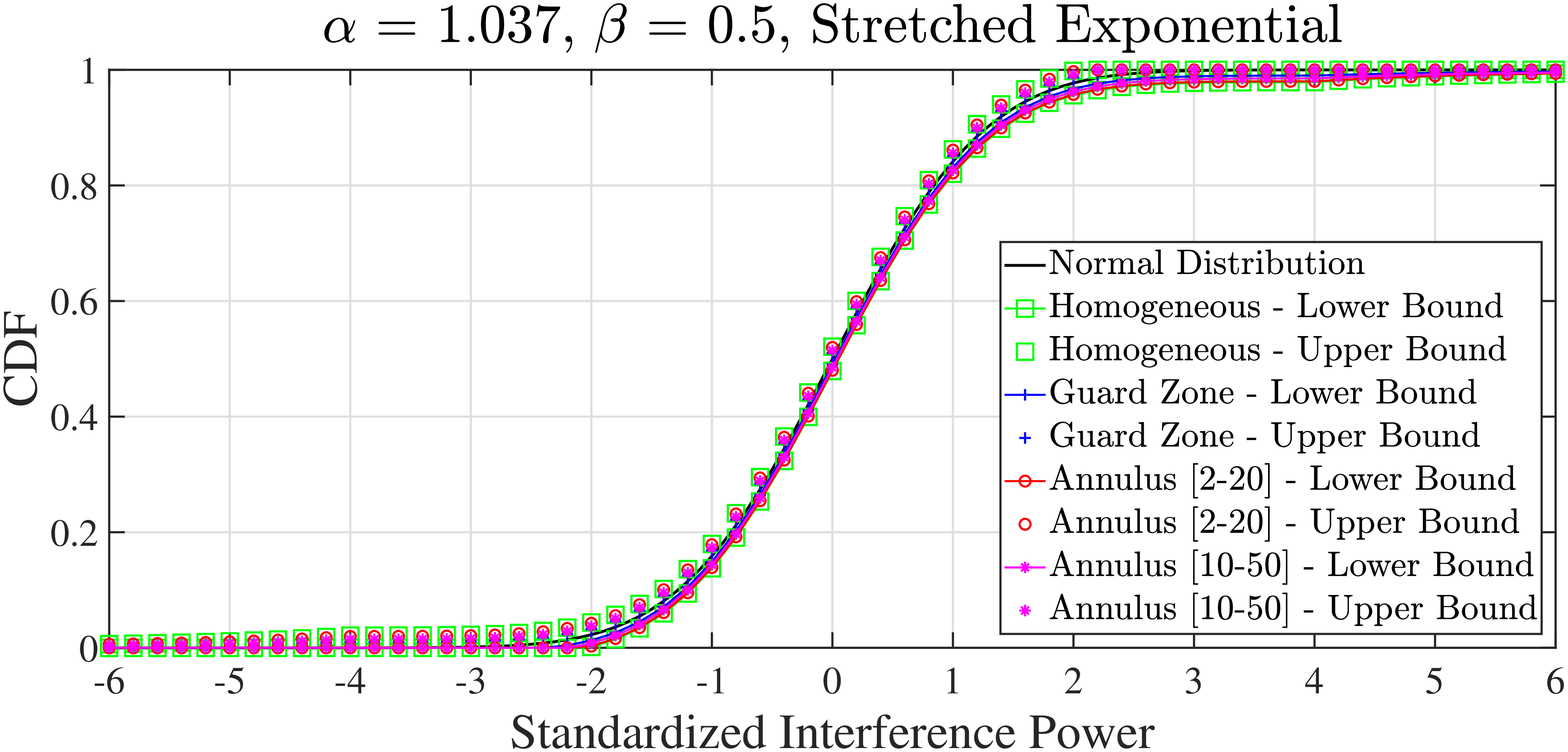}
\end{minipage}
\begin{minipage}[b]{0.45\linewidth} 
\centering
\includegraphics[width=3.35in]{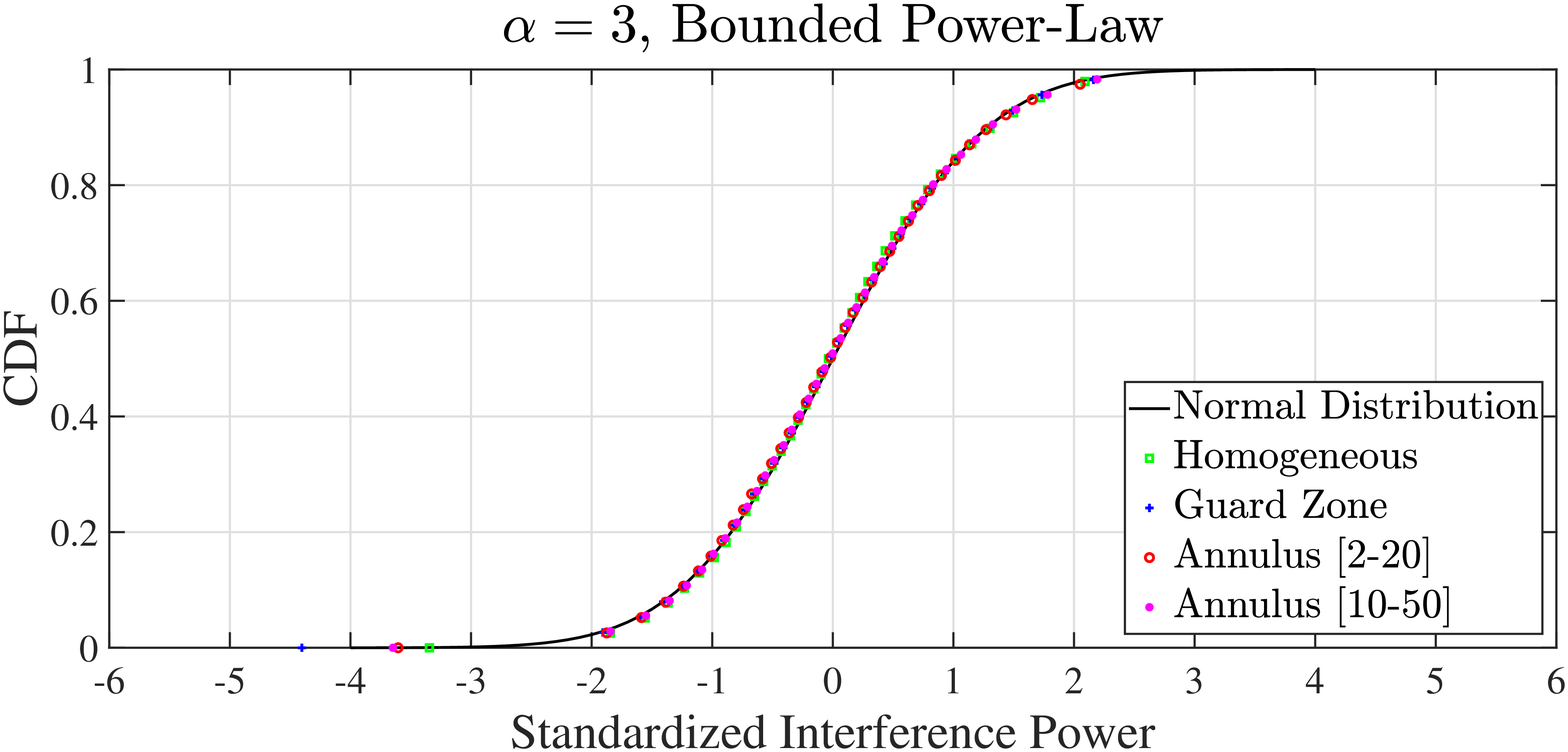}
\end{minipage}
\hspace{0.6cm} 
\begin{minipage}[b]{0.45\linewidth}
\centering
\includegraphics[width=3.35in]{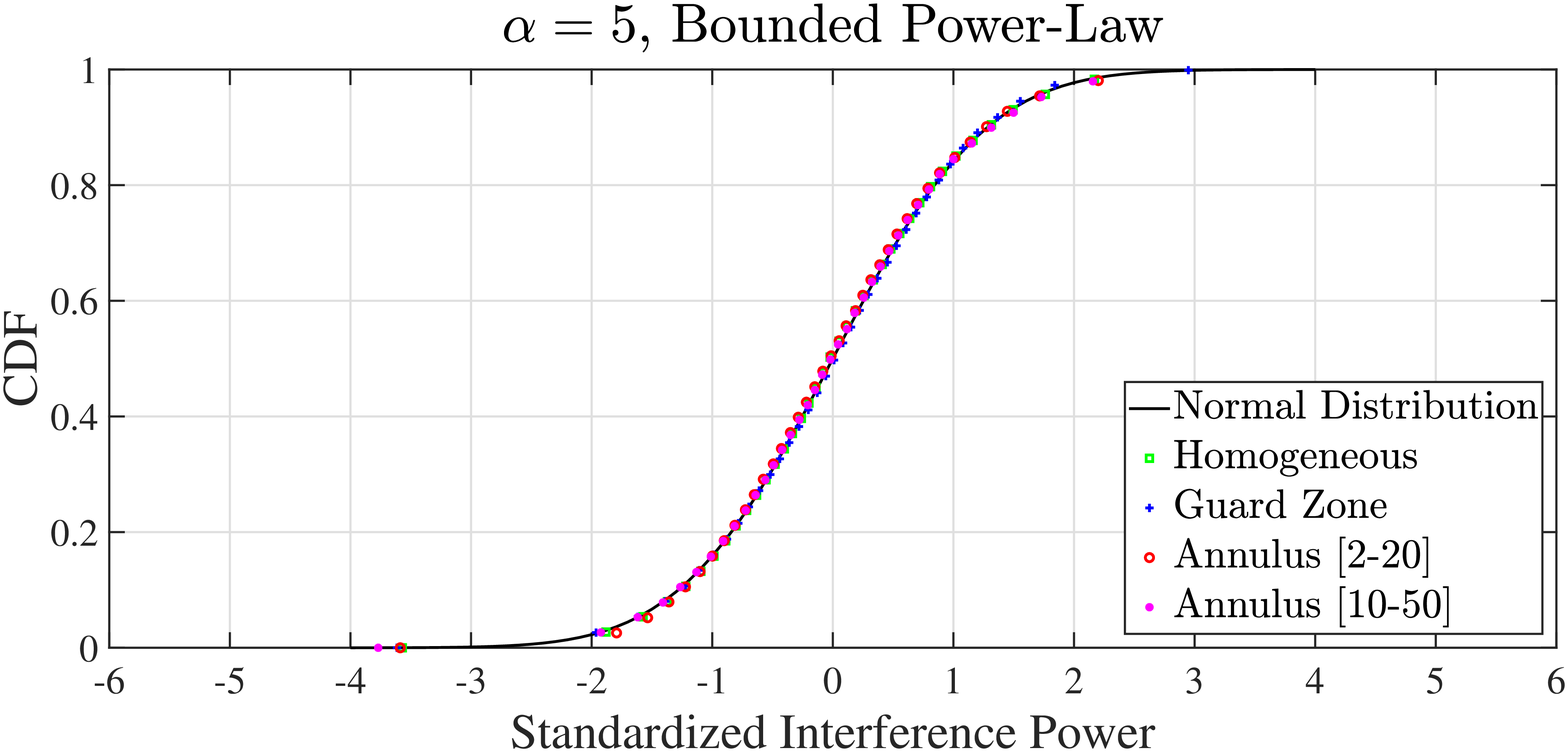}
\end{minipage}
\caption{Gaussian approximation bounds for the standardized AWI CDFs for non-homogeneous BS deployments (top and middle figures). Comparison of the simulated standardized AWI CDFs with the standard normal CDF for non-homogeneous BS deployments (bottom figures). Parameter selection for the stretched exponential distribution is taken from \cite{AlAmmouri17}.  Rayleigh fading with unit mean power is assumed.} 
\label{fig_Non-homogeneous_PPP}
\end{figure*}

We start with the case of {\em homogeneously} distributed BSs over the plane in Fig. \ref{fig1}.  The non-homogeneous case is analyzed by introducing no-BS zones in Fig. \ref{fig_Non-homogeneous_PPP}. We only plot the CDF of AWI in Fig. \ref{fig1} since Theorem \ref{Theorem_1} measures the deviations between the CDFs of standardized AWI and a standard normal random variable.  In the top and middle figures in Fig. \ref{fig1}, we present the upper and lower bounds for the deviations between the standardized AWI and normal distributions, i.e., we plot the expressions $\Psi(x) + \Xi \cdot c(x) $ and $\Psi(x) - \Xi \cdot c(x)$ appearing in Theorem \ref{Theorem_1}, for the two path-loss models with a variety of $\kappa$ values. Two different regimes are apparent in these figures. For the moderate values at which we want to estimate the CDF of standardized AWI, i.e., $\PR{\frac{I_{\vecbold{\lambda}} - \ES{I_{\vecbold{\lambda}}}}{\sqrt{\VS{I_{\vecbold{\lambda}}}}} \leq x}$ with moderate $x$ values, our uniform Berry-Esseen bound, which is $\Xi \cdot 0.4785$, provides better estimations for the AWI distribution. On the other hand, for absolute values larger than $4.0359$ at which we want to estimate the CDF of standardized AWI, i.e.,  $\PR{\frac{I_{\vecbold{\lambda}} - \ES{I_{\vecbold{\lambda}}}}{\sqrt{\VS{I_{\vecbold{\lambda}}}}} \leq x}$ with $\abs{x}$ larger than $4.0359$, our non-uniform Berry-Esseen bound, which is $\Xi \cdot \frac{31.935}{1+\abs{x}^3}$, is tighter. These figures also clearly demonstrate the effect of the BS intensity parameters $\lambda_k$ on our Gaussian approximation bounds. As suggested by Lemma \ref{Lemma: Convergence Rate}, the KS distance between the standardized AWI and normal distributions approaches zero at a rate at least as fast as $\frac{1}{\sqrt{\| \vecbold{\lambda} \|_2}}$. Further, even if all BS intensity parameters are fixed, the distance between the upper and lower bounds in Theorem \ref{Theorem_1} disappears at a rate $O\paren{\abs{x}^{-3}}$ as $\abs{x} \to \infty$ due to the non-uniform bound. Lastly, our bounds appear to be tighter for the stretched exponential path-loss model. This is mainly because of the path-loss model dependent constants appearing in Theorem \ref{Theorem_1}.     

When we compare the top left-hand side and right-hand side figures for the bounded power-law path-loss model in Fig. \ref{fig1}, we observe a better approximation behavior for smaller values of the path-loss exponent $\alpha$.  This is again because of the path-loss model dependent constants appearing in Theorem \ref{Theorem_1}. For this particular choice of the path-loss model and BS distribution over the plane, our approximation results benefit from small values of the path-loss exponent, although the difference between them becomes negligible for moderate to high values of $\kappa$. 

We also performed Monte-Carlo simulations to compare simulated standardized AWI distributions with the normal distribution for $10^4$ random BS configurations. For these simulations, we only used the more conservative bounded power-law path-loss model.  The results are expected to be even better for the stretched exponential path-loss model.  Briefly, the bottom figures in Fig. \ref{fig1} provide further numerical evidence for the Gaussian approximation of AWI in HCNs. Surprisingly, there is a good match between the simulated standardized AWI distribution and the standard normal CDF even for sparsely populated HCNs, i.e., $\kappa = 1$, which is better than what is predicted by our upper and lower Gaussian approximation bounds.


Finally, in addition to the numerical verification of the Gaussian approximation bounds in Theorem \ref{Theorem_1} for homogeneous PPPs, we also performed a similar numerical analysis for HCNs with non-homogeneous BS deployments in Fig. \ref{fig_Non-homogeneous_PPP}. In this figure, we provide numerical results illustrating the derived Gaussian approximation bounds for two different types of non-homogeneous spatial distributions for the locations of BSs. The non-homogeneous BS deployment is obtained by allowing the possibility of BSs not being deployed in some certain regions of the network domain, possibly due to some potential environmental limitations. More specifically, we either consider a guard zone of radius $5$ unit distances around the origin or an annulus with two different values for inner radius, i.e., $2$ and $10$ unit distances, and two different values for outer radius, i.e., $20$ and $50$ unit distances. The BSs in all tiers are assumed to be homogeneously distributed over the plane according to a PPP with $\kappa = 50$, without allowing any BS either inside the guard zone or inside the annulus region. 

As Fig. \ref{fig_Non-homogeneous_PPP} indicates, our Gaussian approximation bounds continue to hold in the considered non-homogeneous BS deployment scenarios when the AWI power is measured at the origin. In particular, the stretched exponential path-loss model leads to tighter bounds than the bounded power-law path-loss model does as in the case of homogeneously distributed BSs.  Further, different BS deployment scenarios seem to have little impact on the bounds for the stretched exponential path-loss model. For the bounded power-law path-loss model, on the other hand, the Gaussian approximation bounds for the AWI distribution becomes tighter than those for the case of homogeneously deployed BSs when there is a guard zone around the origin in which no BSs are allowed to lie. 
Considering the annulus region in which no BSs are allowed to lie for the bounded power-law path-loss model, the Gaussian approximation bounds for the AWI distribution are as good as those obtained for the homogeneously deployed HCN scenario with little impact of the annulus radii on the bounds for the steeper path-loss function. Lastly, the simulation results presented for the bounded power-law path-loss model in the bottom figures in Fig. \ref{fig_Non-homogeneous_PPP} corroborate the Gaussian approximation bounds illustrated in the upper figures.  



In the following section, we will make use of the analytical findings in this section to obtain various performance metrics in $K$-tier HCNs under general settings. In the light of these metrics, we will also gain insights into the impact of network {\em densification} on the outage and rate performance of HCNs 
under specific association policies including both biased and non-biased BS selection strategies.

\section{Outage and Rate Performance of $K$-tier HCNs} \label{Section: HCN Outage Performance}
In this section, we will derive the performance bounds on the HCN capacity metrics, i.e., outage capacity, ergodic capacity and $\ASE$, under two specific association policies: (i) a generic association policy and (ii) biased average received signal strength (BARSS) association policy. However, it should be noted that the analytical approach developed below is general enough for any association policy that preserves the Poisson distribution property for BS locations given the information of $\vecbold{X}^\star$. 
The validity and utility of our analytical results will be numerically illustrated by simulations in Section \ref{Section: Simulation_Results} 



\subsection{Generic Association Policy} \label{Generic_AP}

We start our discussion with the generic association policy. The generic association policy is the policy under which the test user is connected to a BS in tier-$k$ at a (deterministic) distance $r > 0$, and the locations of the rest of the (interfering) BSs in each tier form a non-homogeneous PPP over $\R^2\backslash \mathcal{B}\paren{ \vecbold{x}^{(o)}, d_i}$ with mean measure satisfying the functional form $\Lambda_{i} \circ T^{-1}\paren{\mathcal{S}} = \lambda_i \int_{\mathcal{S}} \mu_i(t) dt$ given this connection information for $i=1, \ldots, K$, where $\mathcal{B}\paren{ \vecbold{x}^{(o)}, d_i}$ is the {\em planar} ball centered at $\vecbold{x}^{(o)}$ with radius $d_i \geq 0$. $\mathcal{B}\paren{ \vecbold{x}^{(o)}, d_i}$ can be thought to signify an {\em exclusion} region around the test user due to operation of the HCN network protocol stack.   

The study of the generic association policy, which may seem a little artificial at first sight, will set the stage for us to analyze both outage and ergodic capacity performance of the BARSS association policy later in this section. The following lemma establishes the Gaussian approximation bounds for the distribution of (standardized) aggregate interference $I_{\vecbold{\lambda}}$ at the test user under the generic association policy by specializing Theorem \ref{Theorem_1} to this case.  
\begin{lem} \label{Lemma: Gauss Approximation for Generic Policy}
Under the generic association policy described above,
$$\abs{\PR{ \frac{I_{\vecbold{\lambda}} - \ES{I_{\vecbold{\lambda}}}}{\sqrt{\VS{I_{\vecbold{\lambda}}}}} \leq x} - \Psi(x)} \leq \Xi \cdot c(x) $$
for all $x \in \R$, where $\Psi(x) = \frac{1}{\sqrt{2\pi}}\int_{-\infty}^x \e{- \frac{t^2}{2}} dt$ which is the standard normal CDF, $c(x) = \min\paren{0.4785, \frac{31.935}{1+\abs{x}^3}}$, and $\Xi  = \sum\nolimits_{i = 1}^K {\frac{{{\lambda _i}P_i^3m_{{H^3}}^{ \paren{ i } }\int_{{d_i}}^\infty  {G_i^3 \paren{ t } {\mu _i} \paren{ t } dt} }}{{{{ \paren{ {\sum\nolimits_{i = 1}^K {{\lambda _i}P_i^2m_{{H^2}}^{ \paren{ i } }\int_{{d_i}}^\infty  {G_i^2 \paren{ t } {\mu _i} \paren{ t } dt} } } } }^{\frac{3}{2}}}}}} $.  
\end{lem}

Since the outage and ergodic capacity metrics given in Definition \ref{Def: Outage Metrics} and Definition \ref{Def: Ergodic Metrics} heavily depend on the level of AWI at the test user, the above Gaussian approximation bound plays a key role to obtain performance upper and lower bounds on the outage and rate performance of HCNs. Below, we start with the outage capacity metric. 
\begin{thm} \label{Thm: Outage Probability Bounds - Generic Association}
Let $\zeta_k\paren{h, \tau, r} =\frac{P_k \paren{\frac{h G_k(r)}{\e{\tau} - 1} - \SNR^{-1}_k}\PG - \ES{I_{\vecbold{\lambda}}}}{\sqrt{\VS{I_{\vecbold{\lambda}}}}}$.  Then, $\PRP{\tau\mbox{-outage}}$ under the generic association policy is bounded above and below as
$$1 - \ES{V_k^+\paren{H_k, \tau, r}} \leq \PRP{\tau\mbox{-outage}} \leq 1 - \ES{V_k^-\paren{H_k, \tau, r}},$$
where $H_k$ is a generic random variable with PDF $q_k$, and the functions $V_k^+$ and $V_k^-$ are given as 
\begin{equation}
V_k^+\paren{h, \tau, r} = \min\brparen{1, \Psi\paren{\zeta_k\paren{h, \tau, r}} + \Xi \cdot c\paren{\zeta_k\paren{h, \tau, r}}} \I{h \geq \frac{\SNR_k^{-1}\paren{\e{\tau} -1}}{G_k(r)}} \label{Eqn: Vk plus}
\end{equation}
and
\begin{equation}
V_k^-\paren{h, \tau, r} = \max\brparen{0, \Psi\paren{\zeta_k\paren{h, \tau, r}} - \Xi \cdot c\paren{\zeta_k\paren{h, \tau, r}}} \I{h \geq \frac{\SNR_k^{-1}\paren{\e{\tau} -1}}{G_k(r)}}. \label{Eqn: Vk minus}
\end{equation}    
\end{thm}
\begin{IEEEproof}
Please see Appendix \ref{Appendix: Outage Probability Bounds - Generic Association}.
\end{IEEEproof} 


Using the bounds on $\PRP{\tau\mbox{-outage}}$, we can bound $C_{\rm o}\paren{\gamma}$ for the generic association policy as below.
\begin{thm} \label{Thm: Outage Capacity Generic Association}
$C_{\rm o}\paren{\gamma}$ under the generic association policy is bounded above and below as
$$ C_{\rm o}\paren{\gamma} \leq \sup\brparen{\tau \geq 0: 1 - \ES{V_k^+\paren{H_k, \tau, r}} \leq \gamma} $$
and
$$ C_{\rm o}\paren{\gamma} \geq \sup\brparen{\tau \geq 0: 1 - \ES{V_k^-\paren{H_k, \tau, r}} \leq \gamma}. $$
\end{thm}
\begin{IEEEproof}
The proof follows from that the upper (lower) bound on the outage probability crosses the target outage probability $\gamma$ earlier (later) than $\PRP{\tau\mbox{-outage}}$ as $\tau$ increases.  
\end{IEEEproof}  

An important high level perspective on the detrimental effects of the network interference on the HCN outage performance can be obtained if we study the outage capacity bounds given in Theorem \ref{Thm: Outage Capacity Generic Association} as a function of $\vecbold{\lambda}$, e.g., for homogeneous PPPs. At each fading state $H_k = h$, it can be shown that the outage capacity scales with the BS intensity parameters according to $\Theta \left( {\left\| \vecbold{\lambda} \right\|_2^{ - 1}} \right)$ 
as $\| \vecbold{\lambda} \|_2$ grows to infinity. 
 This observation is different from the scale-invariance property of $\sinr$ statistics with BS intensity observed in some previous work such as \cite{Dhillon12}, \cite{Jo12}, \cite{Madhu14, Madhu16}. The main reason is that the increase in $I_{\vecbold{\lambda}}$ with denser HCN deployments cannot be counterbalanced by an increase in the received power levels with bounded path-loss models.  That is, the benefits of shorter communication distances are eliminated by an increase in network interference levels in dense HCNs when a bounded path-loss model is used to characterize large-scale wireless propagation losses.  From an HCN design perspective, this result implies that it is imperative to set BS intensities at each tier appropriately for the proper delivery of data services with minimum required QoS to the end users.  

Unlike the outage capacity metric, the ergodic capacity is  more suitable for delay insensitive data traffic and obtained by averaging instantaneous data rates over long time intervals. In the following theorem, the bounds on the ergodic capacity are given, where no channel state information (CSI) is assumed at the transmitter side. 
\begin{thm} \label{Thm: Ergodic Capacity - Generic}
$C_{\rm erg}$ under the generic association policy is bounded above and below as
$$ \int_0^\infty  \ES{V_k^-\paren{H_k, \tau, r}}d\tau \le {C_{\rm erg}} \le \int_0^\infty  {\ES{V_k^+\paren{H_k, \tau, r}}d\tau }. $$
\end{thm}
\begin{IEEEproof}
Please see Appendix \ref{proof_Thm_Ergodic_Capacity_Generic}.
\end{IEEEproof}

\subsection{BARSS Association Policy}
Now, we study the HCN outage and rate performance under the BARSS association policy, in which the test user associates itself to the BS $\vecbold{X}^\star$ given by
$$ \vecbold{X}^\star = \mathop{\arg\max}_{\vecbold{X} \in \Phi} \beta_{\vecbold{X}} P_{\vecbold{X}} G_{\vecbold{X}}\paren{\| \vecbold{X} \|_2}, $$ 
where it is understood that $\beta_{\vecbold{X}} = \beta_k$ if $\vecbold{X} \in \Phi_k$. Consider the event $E_k(r)$ that $A^\star = k$ and $R^\star = r$, i.e., $E_k(r)$ is the event that the test user is associated with a tier-$k$ BS at a distance $r$ under the BARSS association policy. Then, the locations of BSs in tier-$i$ form a non-homogeneous PPP over $\R^2 \backslash \mathcal{B}\paren{ \vecbold{x}^{(o)}, Q_i^{(k)}(r)}$ given the event $E_k(r)$ for $i=1, \ldots, K$, where $Q_i^{(k)}(r) = G_i^{-1}\paren{\frac{\beta_k P_k}{\beta_i P_i} G_k(r)}$, and $G_i^{-1}(y) = \inf\brparen{x \geq 0: G_i(x) = y}$ if $y \in \sqparen{0, G_i(0)}$ and zero otherwise.\footnote{Here, for the simplicity of analysis, we assume that path-loss functions are continuous.} 
This observation puts us back into the generic association policy framework, and the derivation of the bounds for the conditional outage probability/capacity and ergodic capacity on the conditioned event $E_k(r)$ proceeds as before. Averaging over the event $E_k(r)$, we obtain bounds on the unconditional outage probability and capacity metrics. To this end, we need the following lemmas. 

\begin{lem} \label{Lemma: Gauss Approximation for BARSS Policy}
Under the BARSS association policy described above, for all $x \in \R$,
$$\abs{ \PR{ \frac{I_{\vecbold{\lambda}} - \ES{I_{\vecbold{\lambda}}}}{\sqrt{\VS{I_{\vecbold{\lambda}}}}} \leq x \ \big| \ E_k(r)} - \Psi(x)} \leq \Xi_k\paren{r} \cdot c(x), $$
where ${\Xi _k}\left( r \right) =  \sum\nolimits_{i = 1}^K {\frac{{{\lambda _i}P_i^3m_{{H^3}}^{ \paren{ i } }\int_{Q_i^{ \paren{ k } } \paren{ r } }^\infty  {G_i^3\paren{ t } \mu_i\paren{t} dt} }}{{{{ \paren{ {\sum\nolimits_{i = 1}^K {{\lambda _i}P_i^2m_{{H^2}}^{ \paren{ i } }\int_{Q_i^{ \paren{ k } } \paren{ r } }^\infty  {G_i^2\paren{ t } \mu_i\paren{t} dt} } } } }^{\frac{3}{2}}}}}} $ 
, and $c(x)$ and $\Psi(x)$ are as defined in Lemma \ref{Lemma: Gauss Approximation for Generic Policy}. 
\end{lem}

We note that this is almost the same result as in Lemma \ref{Lemma: Gauss Approximation for Generic Policy}, except for a small change in the definition of the constant $\Xi$ to show its dependence on the conditioned event $E_k(r)$.  In order to achieve averaging over the event $E_k(r)$, we need to know the connection probability to a tier-$k$ BS and the conditional PDF of the connection distance given that the test user is associated with a tier-$k$ BS.  To this end, we first obtain the connection probability to a tier-$k$ BS, which we denote by $p_k^\star \defeq \PR{A^\star = k}$, in the next lemma for non-homogeneous PPPs.  

\begin{lem} \label{Lemma: Association Probability for non-homogeneous PPPs}
The probability that the test user is associated with a tier-$k$ BS is 
\begin{eqnarray}
p_k^\star &=& \int_0^\infty \prod_{ \myatop{i=1}{i \neq k}  }^K \exp\paren{-{ \Lambda_i\paren{ \mathcal{B}\paren{ \vecbold{x}^{(o)}, Q_i^{(k)}(u)} } } } f_{R_k}(u) du,  \label{Eqn: Association Probability for non-homogeneous PPPs}
\end{eqnarray}
where $Q_i^{(k)}(u) = G_i^{-1}\paren{\frac{\beta_k P_k}{\beta_i P_i} G_k(u)}$, ${\beta_k}$ is the biasing factor for tier-$k$ BSs and $\Lambda_k\paren{ \cdot }$ is the mean measure of $\Phi_k$, $\mathcal{B}\paren{ \vecbold{x}^{(o)}, Q_i^{(k)}(u)}$ is the ball in $\mathbb{R}^2$ centered at $\vecbold{x}^{(o)}$ with radius $Q_i^{(k)}(u)$, and ${f_{{R_k}}}\left( u \right) = {e^{ - \Lambda_k\paren{ \mathcal{B}\paren{ \vecbold{x}^{(o)}, u } } } }\frac{d}{{du}} \Lambda_k\paren{ \mathcal{B}\paren{ \vecbold{x}^{(o)}, u } } $ is the nearest tier-$k$ BS distance distribution for the test user.
\end{lem}
\begin{IEEEproof}
Please see Appendix \ref{Appendix: Association Probability for non-homogeneous PPPs}.
\end{IEEEproof}

Below, we focus on an important special case in which BS locations follow a homogeneous PPP in each tier, which  leads to Lemma \ref{Lemma: Association Probability} below.

\begin{lem} \label{Lemma: Association Probability}
Let $a_0 = 0$, $a_{K+1} = +\infty$ and $a_i = \frac{\beta_i P_i}{\beta_k P_k} G_i(0)$ for $i \in \brparen{1, \ldots, K} \backslash \brparen{k}$. Let $\pi(i)$ be an enumeration of $a_i$'s in descending order, i.e., $a_{\pi(i)} \geq a_{\pi(i+1)}$ for $i=0, \ldots, K-1$. Let $r_i = G_k^{-1}\paren{a_{\pi(i)}}$ for $i=0, \ldots, K$. Then, $p_k^\star$ is given by 
\begin{equation}
p_k^\star = 2\pi\lambda_k \sum_{j=1}^{K} \int_{r_{j-1}}^{r_j} u \exp\paren{-\pi \paren{\lambda_k u^2 + \sum_{i=1}^{j-1}\lambda_{\pi(i)} \paren{Q_{\pi(i)}^{(k)}\paren{u}}^2}}du. \label{Eqn: Association Probability} 
\end{equation}
\end{lem}
\begin{IEEEproof}
Please see Appendix \ref{Appendix: Association Probability}.
\end{IEEEproof} 

Several important remarks are in order regarding Lemma \ref{Lemma: Association Probability}. The integration in \eqref{Eqn: Association Probability} is with respect to the nearest tier-$k$ BS distance distribution to which the test user is associated. Hence, the BSs in some tiers are inactive with regard to contributing to the association probability for different ranges of the nearest distance from $\Phi_k$ to the origin, which is why we divide the integration limits into disjoint intervals from $r_{j-1}$ to $r_j$ for $j=1, \ldots, K$.  This behavior is different than that observed in \cite{Jo12}, which is again a manifestation of the bounded nature of the path-loss model. In the next lemma, we derive the PDF of $R^\star$ given $A^\star = k$ for a non-homogeneous PPP in each tier.    

\begin{lem} \label{Non-homogeneous_Conditional_Connection_Distance}
The PDF of $R^\star$ given $A^\star = k$ is
\begin{eqnarray}
{f_k}\paren{ u } = \frac{1}{{p_k^ \star }} \prod_{ \myatop{i=1}{i \neq k} }^K \exp\paren{-{ \Lambda_i\paren{ \mathcal{B}\paren{ \vecbold{x}^{(o)}, Q_i^{(k)}(u)} } } } f_{R_k}(u),  \label{Eqn: Non-homogeneous Conditional Connection Distance}
\end{eqnarray}
where ${f_{{R_k}}}\left( u \right)$ is as given in Lemma \ref{Lemma: Association Probability for non-homogeneous PPPs}.
\end{lem}
\begin{IEEEproof}
Please see Appendix \ref{proof_Non-homogeneous_Conditional_Connection_Distance}.
\end{IEEEproof} 

In the next lemma, we again consider the important special case of homogeneous PPPs. 

\begin{lem} \label{Conditional_Connection_Distance}
Let $a_0 = 0$, $a_{K+1} = +\infty$ and $a_i = \frac{\beta_i P_i}{\beta_k P_k} G_i(0)$ for $i \in \brparen{1, \ldots, K} \backslash \brparen{k}$. Let $\pi(i)$ be an enumeration of $a_i$'s in descending order, i.e., $a_{\pi(i)} \geq a_{\pi(i+1)}$ for $i=0, \ldots, K-1$. Let $r_i = G_k^{-1}\paren{a_{\pi(i)}}$ for $i=0, \ldots, K$.  Then, the conditional PDF $f_k(u)$ of $R^\star$ given $A^\star = k$ is given as 

\begin{equation}
f_k(u) = \frac{2\pi\lambda_k}{p_k^\star} \sum_{j=1}^{K} u \exp\paren{-\pi\paren{\lambda_k u^2 + \sum_{i=1}^{j-1} \lambda_{\pi(i)}\paren{Q_{\pi(i)}^{(k)}(u)}^2}} \I{u \in [r_{j-1}, r_j)}. \label{Eqn: Conditional Connection Distance}
\end{equation}
\end{lem}
\begin{IEEEproof}
Please see Appendix \ref{proof_Conditional_Connection_Distance}.
\end{IEEEproof} 

The conditional connection PDF $f_k(u)$ given in \eqref{Eqn: Conditional Connection Distance} can be significantly simplified for small numbers of tiers. 
A reduced expression for one particular but important case of a two-tier HCN is given by the following corollary.  
\begin{cor}\label{Corollary: Two-Tier Connection Distance}
Assume $K=2$, $\beta_1 P_1 G_1(0) \leq \beta_2 P_2 G_2(0)$ and $u^\star = G_2^{-1}\paren{\frac{\beta_1 P_1}{\beta_2 P_2} G_1(0)}$. Then,
$$f_1(u) = \frac{2 \pi \lambda_1}{p_1^\star} u \exp\paren{-\pi\paren{\lambda_1 u^2 + \lambda_2 \paren{Q_2^{(1)}(u)}^2}} \I{u \geq 0} $$
and 
\begin{eqnarray*}
f_2(u) =  \frac{2 \pi \lambda_2}{p_2^\star} u\exp\paren{-\pi \lambda_2 u^2} \I{u < u^\star} + \frac{2 \pi \lambda_2}{p_2^\star} u \exp\paren{-\pi\paren{\lambda_2 u^2 + \lambda_1 \paren{Q_1^{(2)}(u)}^2}} \I{u \geq u^\star}.
\end{eqnarray*}
\end{cor} 


Using these preliminary results, the performance bounds on the outage probability, outage capacity, ergodic capacity and $\ASE$ under the BARSS association policy are given in theorems below. 
\begin{thm} \label{Thm: Outage Probability - BARSS}
Let $\widehat{V}_k^{\pm}\paren{h, \tau, r}$ be defined as in \eqref{Eqn: Vk plus} and \eqref{Eqn: Vk minus}, respectively, by replacing $\Xi$ with $\Xi_k(r)$ given in Lemma \ref{Lemma: Gauss Approximation for BARSS Policy}. Then, $\PR{\tau\mbox{-outage}}$ under the BARSS association policy is bounded below and above as
$$\PR{\tau\mbox{-outage}} \geq 1 - \sum_{k=1}^K p_k^\star \int_0^\infty f_k(r) \ES{\widehat{V}_k^+\paren{H_k, \tau, r}} dr$$
and
$$\PR{\tau\mbox{-outage}} \leq 1 - \sum_{k=1}^K p_k^\star \int_0^\infty f_k(r) \ES{\widehat{V}_k^-\paren{H_k, \tau, r}} dr.$$
\end{thm}
\begin{IEEEproof}
The proof follows from calculating these bounds for $\PR{\tau\mbox{-outage} \ | \ E_k(r)}$ using Theorem \ref{Thm: Outage Probability Bounds - Generic Association}, and then averaging them by using \eqref{Eqn: Association Probability for non-homogeneous PPPs} and \eqref{Eqn: Non-homogeneous Conditional Connection Distance}.
\end{IEEEproof}
\begin{thm} \label{Thm: Outage Capacity - BARSS}
$C_{\rm o}\paren{\gamma}$ under the BARSS association policy is bounded above and below as
$$ C_{\rm o}\paren{\gamma} \leq \sup\brparen{\tau \geq 0: 1 - \sum_{k=1}^K p_k^\star \int_0^\infty f_k(r) \ES{\widehat{V}_k^+\paren{H_k, \tau, r}} dr \leq \gamma}$$
\normalsize
and
$$ C_{\rm o}\paren{\gamma} \geq \sup\brparen{\tau \geq 0: 1 - \sum_{k=1}^K p_k^\star \int_0^\infty f_k(r) \ES{\widehat{V}_k^-\paren{H_k, \tau, r}} dr \leq \gamma}.$$
\normalsize
\end{thm}
\begin{IEEEproof}
The proof follows from that the upper (lower) bound on the outage probability crosses the target outage probability $\gamma$ earlier (later) than $\PRP{\tau\mbox{-outage}}$ as $\tau$ increases.  
\end{IEEEproof}  

Similar to Theorem \ref{Thm: Ergodic Capacity - Generic}, the performance bounds for achievable ergodic capacity are given in the next theorem.
\begin{thm} \label{Thm: Ergodic Capacity - BARSS}
The ergodic capacity achievable under the BARSS association policy is bounded above and below as
$$ { C_{\rm erg} } \le \sum\nolimits_{k = 1}^K {p_k^{\star} \int_0^\infty  {\int_0^\infty  {\EW[ { \widehat{V}_k^ + \paren{ {H,\tau ,r} } } ]} } \,d\tau {f_k}\paren{ r } dr} $$
and
$$ { C_{\rm erg} } \ge \sum\nolimits_{k = 1}^K {p_k^ {\star} \int_0^\infty  {\int_0^\infty  {\EW[ { \widehat{V}_k^- \paren{ {H,\tau ,r} } } ]} } \,d\tau {f_k} \paren{ r } dr,} 
$$
where ${\widehat{V}_k^+\paren{H_k, \tau, r}}$ and ${\widehat{V}_k^-\paren{H_k, \tau, r}}$ are defined as in Theorem \ref{Thm: Outage Probability - BARSS}.  
\end{thm}
\begin{IEEEproof}
The ergodic capacity can be written as $ {C_{\rm erg}} = \sum\nolimits_{k = 1}^K {p_k^ \star \int_0^\infty  {C_{\rm erg}{ \paren{ k, r} }{f_k}{ \paren{ r } }dr} }$, where 
$C_{\rm erg}{ \paren{ k, r } }$ is the conditional ergodic capacity given that the test user is connected to a BS at a distance $r$ and located 
in tier-$k$ under the BARSS association policy. Then, similar to Appendix \ref{proof_Thm_Ergodic_Capacity_Generic}, the proof follows from calculating these bounds for $1 - \PR{\tau\mbox{-outage} \ | \ E_k(r)}$ using Theorem \ref{Thm: Outage Probability Bounds - Generic Association}, and then averaging them by using (\ref{Eqn: Association Probability for non-homogeneous PPPs}) and (\ref{Eqn: Non-homogeneous Conditional Connection Distance}) for non-homogeneous PPPs, or by using \eqref{Eqn: Association Probability} and \eqref{Eqn: Conditional Connection Distance} for homogeneous PPPs, respectively. 
\end{IEEEproof} 


Available spectrum for wireless networks is often very limited. Hence, it is of prime importance to investigate the $\ASE$ of a multi-tier HCN defined as the sum of the achievable bit rates per second per hertz per unit area.  The formal definition is given in Definition \ref{Def: Area Spectral Efficiency}. In the next theorem, we provide upper and lower bounds for the $\ASE$ metric in a $K$-tier HCN, which requires the calculation of conditional outage capacity in each tier. For this calculation, we allow the possibility of having different target outage probabilities for different tiers in order to provide a flexibility of setting a balance between link reliability and rate. The target outage probability for tier-$k$, $k=1, \ldots, K$, will be represented by $\gamma_k \in \paren{0, 1}$ in Theorem \ref{Thm: Area Spectral Efficiency - BARSS}.  We only consider homogeneous PPPs in Theorem \ref{Thm: Area Spectral Efficiency - BARSS} to avoid technical complexities arising in the case of non-homogeneous PPPs. Further, since the $\ASE$ metric characterizes the {\em collective} network performance, rather than the one observed at a specific point as above, it is assumed that all BSs are serving a user.      

\begin{thm} \label{Thm: Area Spectral Efficiency - BARSS}
The $\ASE$ under the BARSS association policy is bounded above and below as
$$ {\ASE}\paren{ {\vecbold{\lambda}},{\vecbold{\gamma}} } \le \sum\nolimits_{k = 1}^K { {\lambda _k} \paren{1 - \gamma_k} { C_{\rm o}^ + \paren{ {k,{\gamma _k}} } } } 
$$
and
$$ {\ASE} \paren{ {\vecbold{\lambda}},{\vecbold{\gamma}} } \ge \sum\nolimits_{k = 1}^K { {\lambda _k}\paren{1 - \gamma_k} { C_{\rm o}^ - \paren{ {k,{\gamma _k}} } } }, 
$$
where $C_{\rm o}^ + \paren{ {k,{\gamma _k}} } \buildrel \Delta \over = \sup \left\{ {\tau  \ge 0:\rho _k^ +  \le {\gamma _k}} \right\}$ and $ C_{\rm o}^ - \paren{ {k,{\gamma _k}} } \buildrel \Delta \over = \sup \left\{ {\tau  \ge 0:\rho _k^ -  \le {\gamma _k}} \right\} $ are conditional outage capacity upper and lower bounds in tier-$k$, $\vecbold{\lambda}  = {\left[ {{\lambda _1}, \ldots ,{\lambda _K}} \right]^{\rm{T}}}$, $\vecbold{\gamma}  = {\left[ {{\gamma _1}, \ldots ,{\gamma _K}} \right]^{\rm{T}}}$, and $\rho _k^ +  \buildrel \Delta \over = 1 - \int_0^\infty  { {f_k}\left( r \right)\ES{\widehat{V}_k^+\paren{H_k, \tau, r}} dr } $ and $ \rho _k^ -  \buildrel \Delta \over = 1 - \int_0^\infty  { {f_k}\left( r \right) \ES{\widehat{V}_k^-\paren{H_k, \tau, r}} dr }$ are conditional outage probabilities given that a mobile user is connected to a tier-$k$ BS, ${\widehat{V}_k^+\paren{H_k, \tau, r}}$ and ${\widehat{V}_k^-\paren{H_k, \tau, r}}$ are given as in Theorem \ref{Thm: Outage Probability - BARSS} specialized to the homogeneous PPPs, and ${f_k}\left( r \right)$ is given in (\ref{Eqn: Conditional Connection Distance}).
\end{thm}
\begin{IEEEproof}
The proof easily follows from Theorems \ref{Thm: Outage Probability - BARSS} and \ref{Thm: Outage Capacity - BARSS}.  
\end{IEEEproof} 

\section{Simulations and Numerical Results}\label{Section: Simulation_Results}
In this part, we present our simulation results illustrating the upper and lower bounds on the HCN capacity metrics derived in Section \ref{Section: HCN Outage Performance}.  We will focus only on the homogeneous PPPs in this section.  The main reason for this choice is to be able to relate the obtained HCN performance metrics to the conclusions drawn in previous papers since homogeneous PPPs are the {\em de facto} BS deployment model in almost all previous work. Further, without having detailed data for the BS deployment, homogeneous PPPs are the most appropriate and widely accepted model for illustrating the derived analytical results.

\begin{figure*}[!t]
\begin{minipage}[b]{0.45\linewidth} 
\centering
\includegraphics[width=3.35in]{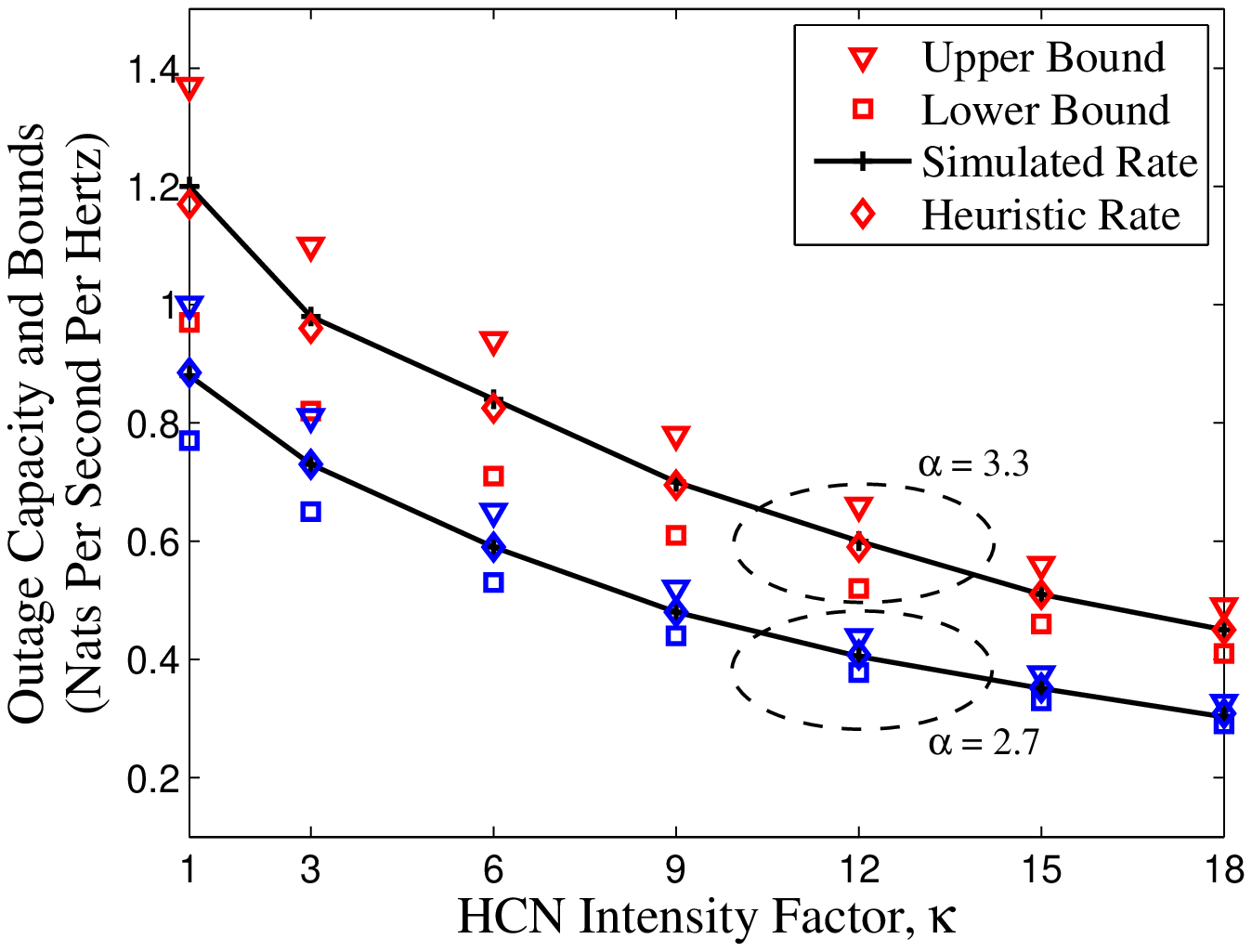}
\end{minipage}
\hspace{0.9cm} 
\begin{minipage}[b]{0.45\linewidth}
\centering
\includegraphics[width=3.35in]{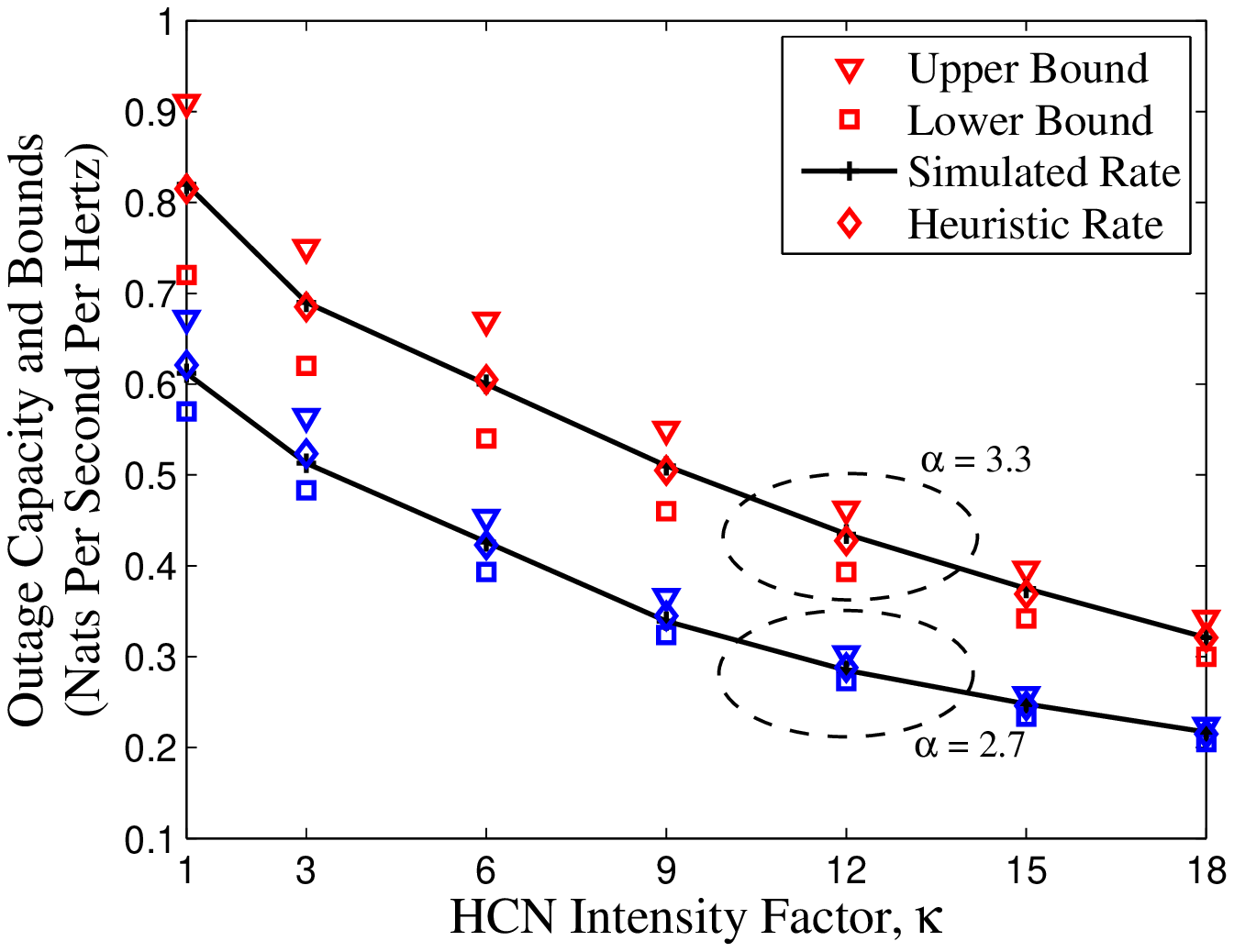}
\end{minipage}
\caption{Upper and lower bounds on ${C_{\rm o}}\left( \gamma  \right)$ for 2- and 3-tier HCNs in the left-hand side and right-hand side figures, respectively.}
\label{fig: outage capacity vs kappa}
\end{figure*}

In particular, we will investigate $C_{\rm o}\paren{\gamma}$, $C_{\rm erg}$ and ${\ASE} \paren{ {\vecbold{\lambda}},{\vecbold{\gamma}} }$ under the BARSS association policy. 
$N_0$ is set to zero and all fading coefficients are independently drawn from Nakagami-$m$ distribution with unit mean power gain and $m = 5$. The path-loss function is conservatively taken to be $G(t) = \frac{1}{1 + t^\alpha}$ for all tiers. Considering the Gaussian approximation bounds illustrated in Fig. \ref{fig1}, the stretched exponential path-loss model is expected to lead to tighter performance bounds. The transmission powers are set as $P_1 = 10 P_2 = 50 P_3$, while we set BS  locations according to homogeneous PPPs with intensities given as $\lambda_1 = 0.1 \kappa$, $\lambda_2 = \kappa$ and $\lambda_3 = 5 \kappa$. Here, $\kappa$ is our (unit-less) control parameter to control the average number of BSs per unit area. For the $2$-tier scenario, only $\brparen{P_k, \lambda_k}_{k=1}^2$ are considered. The target outage probability is $0.15$ for Fig. \ref{fig: outage capacity vs kappa} and $\PG$ is set to $25$ for both figures. For the sake of simplicity, we assume that network layer queues at BSs are fully-loaded, an extension of which to the lightly loaded case is elaborated in Section \ref{Section: Extentions} and will be considered as a future work \cite{Dhillon13a}.\footnote{Our parameter selections in this section are deliberately different than those in Section \ref{Section: Gaussian_Approximation} in order to illustrate that our results continue to hold broadly for any set of network parameter combinations.}

We plot the bounds in Theorem \ref{Thm: Outage Capacity - BARSS} on ${C_{\rm o}}\left( \gamma  \right)$ for $2$-tier and $3$-tier HCNs as a function of $\kappa$ in Fig. \ref{fig: outage capacity vs kappa}. Two different values of $\alpha$ are used. As this figure shows, both upper and lower bounds approximate $C_{\rm o}\paren{\gamma}$ within $0.06$ Nats$/$Sec$/$Hz for $\alpha = 2.7$ and within $0.15$ Nats$/$Sec$/$Hz for $\alpha = 3.3$ in the $2$-tier scenario. They are tighter for the $3$-tier scenario due to denser HCN deployment. The heuristic rate curve, which is the arithmetic average of the upper and lower bounds, almost perfectly track $C_{\rm o}\paren{\gamma}$ for all cases considered in Fig. \ref{fig: outage capacity vs kappa}.       

An interesting observation is the monotonically decreasing nature of $C_{\rm o}\paren{\gamma}$ with $\kappa$. This is in accordance with the discussion on the $\Theta \left( {\left\| \vecbold{\lambda} \right\|_2^{ - 1}} \right)$-type 
scaling behavior of outage capacity in Section \ref{Section: HCN Outage Performance}. Hence, we cannot improve the downlink data rates indefinitely in an HCN by adding more BS infrastructure.  We must either mitigate interference more efficiently or find the optimum BS intensity per tier maximizing delivered data rates per unit area.



\begin{figure}
\begin{minipage}[t]{8cm}
\centering{\includegraphics[width=3.35in]{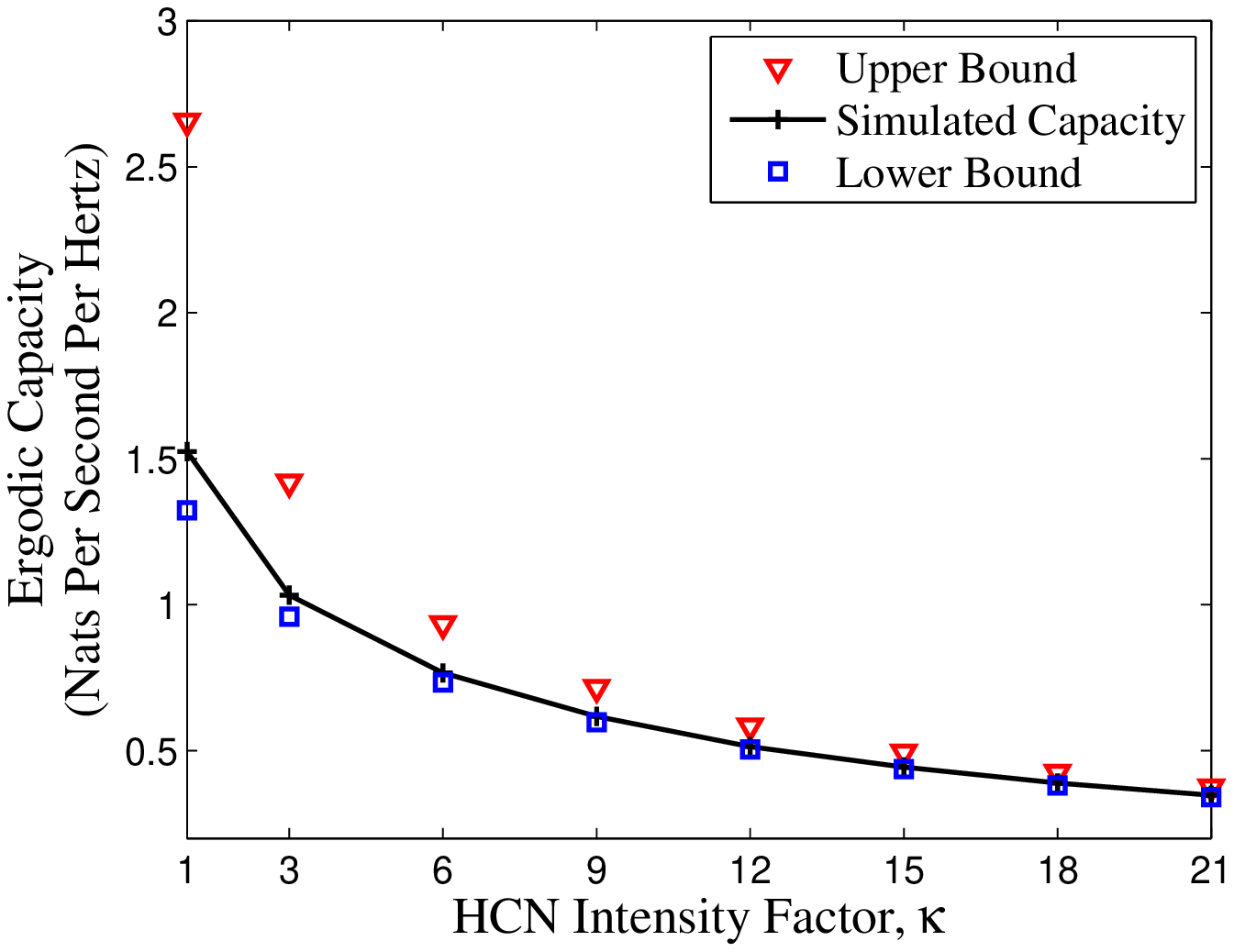}}
\caption{Change of ergodic capacity for various values of $\kappa$. $\lambda_1 = 0.1 \kappa$, $\lambda_2 = \kappa$. $\alpha$, $\beta_{1}$ and $\beta_2$ are set to 3, 1 and 1, respectively. }
\label{fig: Ergodic vs kappa}
\end{minipage}
\hfill
\begin{minipage}[t]{8cm}
\centering{\includegraphics[width=3.35in]{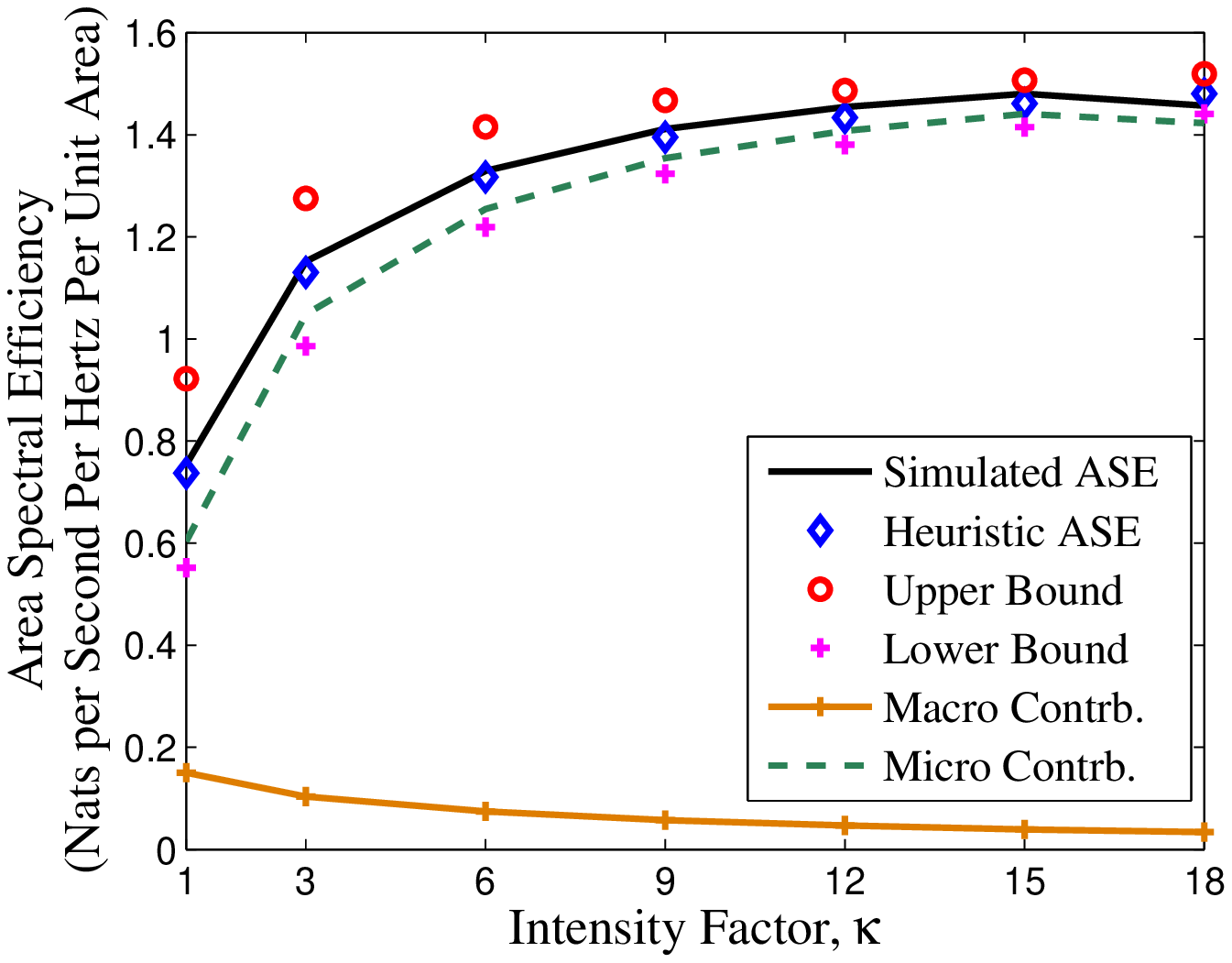}}
\caption{Change of area spectral efficiency for various values of $\kappa$. $\lambda_1 = 0.1$, $\lambda_2 = \kappa$, $\beta_{\rm ratio} = \beta_2 / \beta_1 = 3$ and $\alpha = 3$. }
\label{fig: ASE vs kappa }
\end{minipage}
\end{figure}

We plot $C_{\rm erg}$ and the corresponding bounds given in Theorem \ref{Thm: Ergodic Capacity - BARSS} for the 2-tier HCN scenario as a function of ${\kappa}$ in Fig. \ref{fig: Ergodic vs kappa}. We set the path-loss exponent ${\alpha}$ to 3. For moderate values of ${\kappa}$, we observe that while our upper bound approximates $C_{\rm erg}$ within {\em one} Nats$/$Sec$/$Hz, the lower bound is closer than $0.3$ Nats$/$Sec$/$Hz. Our bounds become very close to the simulated rate for moderate to high values of ${\kappa}$, and especially the gaps among the bounds and the $C_{\rm erg}$ become negligibly small for high values of ${\kappa}$.


When we compare the bounds on outage and ergodic capacities depicted in Figs. \ref{fig: outage capacity vs kappa} and \ref{fig: Ergodic vs kappa}, an important observation we have is the upper bound for the ergodic capacity being looser than the one for the outage capacity in sparse HCNs. The main reason for this behavior is the nature of Gaussian approximation bounds appearing in Theorem \ref{Theorem_1}.  In particular, the outage probability, and hence the outage capacity, depends on the tail of the AWI distribution. Due to the existence of the function $c(x)$ in the Gaussian approximation bounds in Theorem 1, the derived bounds on the tails of the standardized downlink AWI distribution are tight even for the {\em sparsely} deployed HCNs. The ergodic capacity calculations, on the other hand, require an averaging over the entire range of AWI values and this range includes the interference values for which our bounds are not as tight as for higher interference values. This leads to the observed behavior of obtained bounds being a little looser for the ergodic capacity, especially the upper bound, when compared to those obtained for outage capacity.

Finally, Fig. \ref{fig: ASE vs kappa } depicts the changes of $\ASE$ for a $2$-tier HCN scenario as a function of $\kappa$, where BS intensities are set as $\lambda_1 = 0.1$ and $\lambda_2 = \kappa$.  Note that we fix $\lambda_1$ at $0.1$ to see how $\ASE$ changes by adding more unplanned infrastructure to the network as $\lambda_2$ grows. Biasing coefficient for tier-$1$, i.e., $\beta_1$, is assumed to be unity while the biasing value $\beta_2$ for tier-$2$ is set to $3$. Target outage probabilities in both tiers are assumed to be identical, {\em i.e.,} $\gamma_k = \gamma$, and taken as $0.15$ for maintaining the same link reliability in each tier.  

We firstly observe that our upper and lower bounds again accurately characterize the simulated $\ASE$ curve. Further, the heuristic $\ASE$ curve, which is taken to be the arithmetic average of the upper and lower bounds, almost perfectly tracks the simulated $\ASE$.  We note three different forms of behavior of $\ASE$ as a function of $\kappa$. For sparse to moderate values of ${\kappa}$, {i.e.}, between $1$ and $6$, $\ASE$ increases rapidly due to efficient utilization of the spectrum which is made available by tier-$2$ BSs. For $\kappa$ from moderate to high values, i.e., between $6$ and $12$, improvement in $\ASE$ starts to slow down due to the pressure of growing AWI. For ${\kappa}$ values higher than $12$, $\ASE$ almost stops improving due to the overwhelming growth in AWI.  Especially, the contribution of macro BSs to the overall $\ASE$ is much more negatively affected than that of micro BSs by the elevated AWI levels in the dense HCN regime, as illustrated by the macro and micro contribution curves. These observations point out that significant $\ASE$ gains can be achievable through biasing along with applying an efficient interference management and suppression in hyper-dense multi-tier HCNs. Otherwise, the network density should be carefully adjusted to maintain target $\ASE$.

\section{Discussion: Scope of the Paper, Extensions and Limitations of the Proposed Approach} \label{Section: Extentions}
Our main aim in this paper is to provide tight Gaussian approximation bounds for AWI in HCNs and to illustrate the utility of these bounds for the system-level HCN performance analysis. Hence, we consider a simple but practically relevant system model in Section \ref{System_Model} for the sake of avoiding more involved analysis in Section \ref{Section: Gaussian_Approximation} and Section \ref{Section: HCN Outage Performance}. In this section, on the other hand, we discuss some notable extensions of our baseline model and the required changes for these extensions.

\subsection{Load-Aware Modeling and Analysis}
Our network model in Section \ref{System_Model} assumes that all the BSs are fully-loaded and have access to the same communication resources both in time and frequency. However, in reality, small cells have fewer users with lighter traffic loads \cite{Dhillon13a}. This observation implies that the fully-loaded network scenario can only provide lower bounds for the actual network performance. Hence, an important extension of our results in this paper lies in the design and analysis of load-aware HCNs \cite{Dhillon13a, Renzo16, Shojaeifard14}. 

In particular, the load-aware analysis of HCNs depends primarily on the thinning of the underlying BS location processes, which can be categorized into two classes of (i) uncorrelated thinning \cite{Dhillon13a,Renzo16} and (ii) correlated thinning \cite{Shojaeifard14}. For the case of uncorrelated thinning, it is important not to turn off the serving BS and the analysis of the total received power at a typical point plays an important role to discover the outage probability in load-aware HCNs \cite{Dhillon13a}. In this regard, our main Gaussian approximation result in Theorem \ref{Theorem_1} can be directly applied to the conditionally thinned BS location processes, without turning off the serving BS, in order to obtain the outage probability and other related performance measures. Our framework is flexible enough to allow the consideration of both location independent thinning \cite{Dhillon13a} and location dependent thinning \cite{Renzo16} in a similar fashion. 

For the case of correlated thinning, on the other hand, the thinning probability depends on the configuration of all BSs \cite{Shojaeifard14}. Therefore, it is not clear if the thinned BS location processes will still maintain their Poisson property. If they do, Theorem \ref{Theorem_1} can still be applied to approximate the AWI distribution. Otherwise, our approach in this paper cannot be used since the Poisson assumption is critical for the proof of Theorem \ref{Theorem_1}. 

\subsection{Multiple-Input Multiple-Output Communications}
In addition to network densification, another communication strategy that features prominently in IEEE 802.11e WiMAX and 3GPP LTE-A standards is multiple-input multiple-output (MIMO) communications \cite{Renzo_Lu15, SIE16, Gupta15, Dhillon13b, Shojaeifard16}. Multiple antenna techniques offer both multiplexing and diversity gains in order to improve spectral efficiency \cite{Tse05}. Two important modifications are needed in order to accommodate MIMO techniques in our baseline model in Section \ref{System_Model}. The first is the consideration of {\em beamforming}, which will bring additional degrees-of-freedom for communications over single-input single-output (SISO) systems. The second is the {\em inter-antenna interference}. 

More specifically, assuming that a BS belonging to tier-$k$ has $M_{k}$ transmission antennas and serves $N_k$ single-antenna mobile users,\footnote{Usually, $N_k$ is taken to be smaller than $M_k$.} the transmitted signal from a BS located at $\vec{X} \in \Phi_k$ can be represented as 
\begin{eqnarray*}
\vec{s}_{\vec{X}} = \sqrt{P_k} \sum_{i=1}^{N_k} \vec{b}_{\vec{X}, i} s_{\vec{X}, i}, 
\end{eqnarray*}
where $\brparen{\vec{b}_{\vec{X}, i}}_{i=1}^{N_k}$ and $\brparen{s_{\vec{X}, i}}_{i=1}^{N_k}$ are the sets of beamforming vectors and data symbols used by the BS $\vec{X} \in \Phi_k$ to communicate with $N_k$ mobile users simultaneously. Then, a test user $o$ located at $\vec{x}^{(o)}$ and connected to the BS $\vec{X}^\star \in \Phi_k$ will receive the following baseband signal       
\begin{eqnarray*}
\lefteqn{y_{o} = \underbrace{\sqrt{P_k G_k\paren{T\paren{\vec{X}^\star}}}  \vec{h}_{\vec{X}^\star}^\top \vec{b}_{\vec{X}^\star, o} s_{\vec{X}^\star, o}}_{\mbox{intended signal}} + \underbrace{\sqrt{P_k G_k\paren{T\paren{\vec{X}^\star}}} \sum_{i=1, i \neq o}^{N_k} \vec{h}_{\vec{X}^\star}^\top \vec{b}_{\vec{X}^\star, i} s_{\vec{X}^\star, i}}_{\mbox{inter-antenna interference}}} \hspace{15cm} \\
\lefteqn{+ \underbrace{\sum_{\vec{X} \in \Phi \setminus \brparen{\vec{X}^\star}}\sqrt{P_{\vec{X}} G_{\vec{X}}\paren{T\paren{\vec{X}}}} \vec{h}_{\vec{X}}^\top\vec{s}_{\vec{X}}}_{\mbox{inter-cell interference}},} \hspace{14cm}   
\end{eqnarray*} 
where $\vec{h}_{\vec{X}}$ represents the {\em amplitude} fading coefficients for the channel from a BS located at $\vec{X} \in \Phi$ to the test user. As it is clear from this baseband signal representation, the inter-antenna interference is not related to interfering BS locations and can be handled directly by the standard techniques from probability theory. Further, if zero-forcing beamforming is employed, it can be eliminated from $y_o$. The inter-cell interference, on the other hand, depends on the locations of interfering BSs.  However, its structure is the same with the one studied in Section \ref{System_Model} and Section \ref{Section: Gaussian_Approximation}, except the new terms appearing in the form of $\vec{h}_{\vec{X}}^\top \vec{s}_{\vec{X}}$. Hence, the distribution of the interference {\em power} can be approximated exactly as in Theorem \ref{Theorem_1} once the second and third moments for $\abs{\vec{h}_{\vec{X}}^\top \vec{s}_{\vec{X}}}^2$ are determined. Due to the linearity of expectation operator, this is usually a straightforward task. 

Lastly, we only considered single-antenna mobile users above for simplicity. Similar arguments with receiver beamforming not depending on the interfering BS configuration hold when mobile users have multiple antennas. If the receiver beamforming vectors depend on the locations of intefering BSs, then this will induce a correlation between the BS location processes and the random variables modulating the path-loss coefficients. In this case, our Gaussian approximation results cannot be applied directly.   
 
\subsection{Power Control}
Another important and pretty much standard technique in wireless communications is the control of transmission powers through a power control mechanism \cite{Chiang08, Inaltekin12b, ElSawy17}. In particular, transmission powers need to be allocated to each link for data communications, but this allocation has knock-on effects on other links in the network due to the broadcast nature of the shared wireless medium.  Hence, an efficient power control mechanism is critical for achieving high data rates in HCNs. In this paper, for the simplicity of exposition, we do not explicitly consider any power control mechanism in our baseline model in Section \ref{System_Model}, which is also the case in most previous papers \cite{AlAmmouri17, Andrews11, Dhillon12, Jo12, SIE14a, Dhillon13a, Renzo16, Shojaeifard14, Dhillon13b, Shojaeifard16}. However, it is straightforward to incorporate the effect of any pairwise power control mechanism between a transmitter and its receiver on our Gaussian approximation result in Theorem \ref{Theorem_1}. 

To see this more clearly, assume that $P_{\vec{X}}\paren{R_{\vec{X}}, U_{\vec{X}}}$ is the power control mechanism used by the BS located at $\vec{X} \in \Phi$, where $R_{\vec{X}}$ and $U_{\vec{X}}$ are the distance and the (power) fading coefficient between this BS and its receiver. Then, the total interference power at the test user can be written as
\begin{eqnarray*}
I_{\vec{\lambda}} = \sum_{\vec{X} \in \Phi \setminus \brparen{\vec{X^\star}}} P_{\vec{X}}\paren{R_{\vec{X}}, U_{\vec{X}}} H_{\vec{X}} G_{\vec{X}}\paren{T\paren{\vec{X}}}, 
\end{eqnarray*}
which is the same interference power expression in \eqref{Eqn: Total Interference Power}, except the fact that the transmission powers now depend on connection distances between BSs and their intended receivers as well as the corresponding fading gains. We note that the functional form for $P_{\vec{X}}\paren{R_{\vec{X}}, U_{\vec{X}}}$ can be assumed to be the same for all BSs belonging to the same tier.  As a result, using $P_{\vec{X}}\paren{R_{\vec{X}}, U_{\vec{X}}} H_{\vec{X}}$ as marks of BSs, observing that these marks are independent of other BS locations in $\Phi \setminus \brparen{\vec{X}}$ and replacing $P_k^2 m_{H^2}^{(k)}$ and $P_k^3 m_{H^3}^{(k)}$ with the second and third moments of these marks, we can obtain a version of Theorem \ref{Theorem_1} suitable for characterizing AWI distribution in power controlled HCNs. We note that if the employed power control mechanism in a network tier also depends on the configuration of BSs in the network, then the marks are not independent of the BS locations anymore and our Gaussian approximation results cannot be applied directly in this case.        

\subsection{Uplink Communications and Other Network Technologies}
Last but not least, we conclude this section of the paper with a brief discussion on the applicability of our results for the HCN uplink communications and other network technologies. Although our main focus in this paper is on the downlink communications in an HCN setting \cite{Dhillon12}, the same analysis directly extends to the HCN uplink communications \cite{Renzo16b, Yazdi14, Novlan13}. 

To see this, first consider the simplest case of mobile terminals with identical transmission technology without any power control. This is exactly the case of single-tier cellular networks with the understanding of interference signals emanating from mobile users rather than BSs. Hence, the structure of the AWI power at any test BS remains the same as in \eqref{Eqn: Total Interference Power} and Theorem \ref{Theorem_1} can be directly applied to approximate uplink AWI statistics. If the transmission technologies of mobile users are, on the other hand, dissimilar and allow them to use different but constant transmission power levels, perhaps due to different manufacturer specifications, then this case can be treated exactly as in the HCN downlink communications after a classification of transmission technologies into multiple groups. Finally, if a pairwise power control mechanism is employed in the uplink communications, then we can use selected power levels and fading coefficients as marks associated with mobile users and still use Theorem \ref{Theorem_1} as explained above to characterize the uplink AWI statistics in HCNs.

In addition to uplink communications, the analytical framework developed in this paper can find further applications in other emerging network technologies such as Internet-of-Things (IoT) and machine-type communications \cite{Fuqaha15, Sarikaya17}. Alongside mobile-edge computing and fog computing technologies \cite{Chiang16, Taleb17}, it is expected that billions of smart devices will be accessing the same spectrum for Internet connectivity in near future \cite{IDC14}. LTE for machine-type communications (LTE-M), extended coverage GSM (EC-GSM) and narrowband IoT (NB-IoT) are the recent 3GPP study items in order to accommodate billions of IoT devices in the same spectrum via the cellular infrastructure \cite{Ericsson16, Nokia17, Shirvanimoghaddam17}. In addition, LoRa and SigFox have been proposed for the unlicensed spectrum \cite{Raza17}. Due to large numbers of IoT devices, we expect that the Gaussian approximation regime discovered in this paper will continue to hold, possibly much more tightly, in both licensed and unlicensed bands for machine-type communications. Since our network model is built upon general assumptions for path-loss, fading and transmitter location processes, the specific nature of transmitters is irrelevant and Theorem \ref{Theorem_1} can still be applied to obtain Gaussian approximation bounds for the emerging machine-type communications technology once the device locations are characterized through an appropriate PPP. Further details are left as a future work.

\section{Conclusions and Future Work}\label{Conclusion}
This paper has examined various performance metrics of interest for the downlink in dense $K$-tier HCNs under general settings including non-homogeneous PPPs by introducing a principled methodology. 
 To this end, we have first investigated the Gaussian approximation for the downlink AWI distribution in dense $K$-tier HCNs. Analytical bounds measuring the Kolmogorov-Smirnov distance between the AWI distribution and Gaussian distribution have been obtained. The derived bounds have also been illustrated numerically through simulations of particular three-tier HCN scenarios. A good statistical fit between the simulated (centralized and normalized) AWI distribution and the standard normal distribution has been observed even for moderate values of BS intensities. 

Secondly, we have examined the downlink capacity metrics of dense $K$-tier HCNs under general signal propagation models, allowing for the use of general bounded path-loss functions, arbitrary fading distributions and general PPPs. Tight upper and lower bounds on the outage capacity, ergodic capacity and area spectral efficiency have been obtained for two specific association policies - the generic association policy and the BARSS association policy. The validity of our analytical results has also been confirmed by simulations. An important design insight for dense $K$-tier HCNs arising from this study is that the celebrated $\sir$ invariance property does not hold anymore if a bounded path-loss model is used to characterize large-scale wireless propagation losses in the downlink of a $K$-tier HCN.  Hence, it is imperative to either mitigate interference more efficiently or find the optimum BS intensity per tier maximizing delivered data rates to mobile users in order to reap the benefits from network densification. 

The proposed approach can be extended to load-aware analysis of HCNs, MIMO communications, power-controlled transmissions, uplink communications, emerging telecommunications technologies with densely deployed transmitters and other association policies. It has the potential of understanding the dense HCN performance and design beyond specific selections of the path-loss model and the fading distribution. Utilizing these results, the future plans of the authors include the development of novel spectrum management techniques for dense $K$-tier HCNs, an investigation of multi-slope path-loss models for emerging millimeter wave communications under general settings, and an investigation of provably near-optimum network control mechanisms such as optimum hybrid-access and cell-range expansion.

\appendices

\section{ The Proof of Theorem \ref{Thm: Outage Probability Bounds - Generic Association} } \label{Appendix: Outage Probability Bounds - Generic Association}
In this appendix, we will provide the proof for Theorem \ref{Thm: Outage Probability Bounds - Generic Association} establishing the outage capacity bounds for the generic association policy. Given that the test user is associated to a BS at a distance $r$ in tier-$k$, we can express the $\tau$-outage probability as 
\begin{eqnarray*}
\PRP{\tau\mbox{-outage}}  &=& \PR{\log\paren{1 + \sinr_{\mathcal{A}} } < \tau}  \\
&=& 1 - \int_{\frac{\snr_k^{-1}\paren{\e{\tau} - 1}}{G_k(r)}}^\infty \PR{I_{\vec{\lambda}} \leq P_k\paren{\frac{h G_k(r)}{\e{\tau} - 1} - \snr_k^{-1}}\PG} q_k(h) dh, 
\end{eqnarray*}
where the last equality follows from the fact that $I_{\vec{\lambda}}$ is a positive random variable, and we have $P_k \paren{{\frac{{hG_{k}\left( r \right)}}{{{\e{\tau}} - 1}} - {\snr}_k^{ - 1}}} \PG < 0$ if and only if $h < \frac{{\left( {{\e{\tau} } - 1} \right){\snr}_k^{ - 1}}} {{G_{k}\left( r \right)}}$. By using Lemma \ref{Lemma: Gauss Approximation for Generic Policy} and the natural bounds $0$ and $1$ on the probability, we can upper and lower bound $\PRP{\tau\mbox{-outage}}$ as   
\begin{eqnarray*}
\PRP{\tau\mbox{-outage}} &\leq& 1 - \int_{\frac{\snr_k^{-1}\paren{\e{\tau} - 1}}{G_k(r)}}^\infty \max\brparen{0, \Psi\paren{\zeta_k\paren{h, \tau, r}} - \Xi \cdot c\paren{\zeta_k\paren{h, \tau, r}}} q_k(h) dh \\
&=& 1 - \ES{\max\brparen{0, \Psi\paren{\zeta_k\paren{H_k, \tau, r}} - \Xi \cdot c\paren{\zeta_k\paren{H_k, \tau, r}}} \I{H_k \geq \frac{\SNR_k^{-1}\paren{\e{\tau} -1}}{G_k(r)}}} \\
&=& 1 - \ES{V_k^{-}\paren{H_k, \tau, r}}
\end{eqnarray*}
and
\begin{eqnarray*}
\PRP{\tau\mbox{-outage}} &\geq& 1 - \int_{\frac{\snr_k^{-1}\paren{\e{\tau} - 1}}{G_k(r)}}^\infty \min\brparen{1, \Psi\paren{\zeta_k\paren{h, \tau, r}} + \Xi \cdot c\paren{\zeta_k\paren{h, \tau, r}}} q_k(h) dh \\
&=& 1 - \ES{\min\brparen{1, \Psi\paren{\zeta_k\paren{H_k, \tau, r}} + \Xi \cdot c\paren{\zeta_k\paren{H_k, \tau, r}}} \I{H_k \geq \frac{\SNR_k^{-1}\paren{\e{\tau} -1}}{G_k(r)}}} \\
&=& 1 - \ES{V_k^{+}\paren{H_k, \tau, r}},
\end{eqnarray*}
where $\Xi$ and $c(x)$ are as given in Lemma \ref{Lemma: Gauss Approximation for Generic Policy}, $\Psi(x)$ is the standard normal CDF and $\I{\cdot}$ is the indicator function.

\section{The Proof of Theorem \ref{Thm: Ergodic Capacity - Generic} } \label{proof_Thm_Ergodic_Capacity_Generic}
The achievable ergodic capacity can be simply written as
$$
{C_{\rm erg}} = \EW_{{\sinr}_{\mathcal A}}\left[ {\log \paren{ {1 + {{\sinr}_{\mathcal A}}} } } \right] \mathop  = \limits^{\left( a \right)} \int_0^\infty {1 - \PRP{\tau\mbox{-outage}} }  d\tau , \nonumber
$$
where (a) follows from the fact that $\log\paren{1 + \SINR_{\mathcal A}}$ is a positive random variable and hence its expectation is equal to $\int_0^\infty \PR{\log\paren{1 + \SINR_{\mathcal A}} > \tau} d\tau$. Using this observation, we can easily obtain the upper and lower bounds on the ergodic capacity as 
$
{C_{\rm erg}} \le \int_0^\infty  { \EW{ \left[ {V_k^ + \paren{ {{H_k},\tau, r } } } \right] }d\tau } 
$
and
$
{C_{\rm erg}} \ge \int_0^\infty  { \EW{ \left[ {V_k^ - \paren{ {{H_k},\tau, r } } }  \right] }d\tau } 
$
by using the steps similar to those in Appendix \ref{Appendix: Outage Probability Bounds - Generic Association}. 

\section{The Proof of Lemma \ref{Lemma: Association Probability for non-homogeneous PPPs} } \label{Appendix: Association Probability for non-homogeneous PPPs}
In this appendix, we will derive the connection probability of the test user to a serving BS in tier-$k$ under the BARSS association policy, which is denoted by $ p_k^\star {\buildrel \Delta \over =} \PR{A^\star = k}$.  Let $R_i$ be the nearest distance from $\Phi_i$ to the test user for $i=1, \ldots, K$.  Then, utilizing the structure of the BARSS association policy, this probability can be written as
\begin{eqnarray}
p_k^\star 
&\stackrel{\rm (a)}{=}& \int_0^\infty \prod_{ \myatop{i=1}{i \neq k} }^K \PR{\beta_i P_i G_i\paren{R_i} \leq \beta_k P_k G_k\paren{u} \Big| R_k = u} f_{R_k}(u) du \nonumber \\
&\stackrel{\rm (b)}{=}& \int_0^\infty \prod_{ \myatop{i=1}{i \neq k} }^K \PR{\beta_i P_i G_i\paren{R_i} \leq \beta_k P_k G_k\paren{u}} f_{R_k}(u) du, \label{Eqn: Non-homogeneous Connection Prob 1}
\end{eqnarray}
where the identity (a) follows from the conditional independence of the events 
$$\brparen{\beta_i P_i G_i\paren{R_i} \leq \beta_k P_k G_k\paren{R_k}} \hspace{0.15cm} \mbox{ for } \hspace{0.15cm} i \in \brparen{1, \ldots, K} \backslash \brparen{k} $$
given any particular realization of $R_k$, and the identity (b) follows from the independence of the nearest BS distances from different tiers.  Each probability term in \eqref{Eqn: Non-homogeneous Connection Prob 1} can be written as
\begin{eqnarray}
\PR{\beta_i P_i G_i\paren{R_i} \leq \beta_k P_k G_k\paren{u}} 
= \PR{R_i \geq Q_i^{(k)}(u)}
= \exp\paren{-{ \Lambda_i\paren{ \mathcal{B}\paren{ \vecbold{ X^{(0)} }, Q_i^{(k)}(u)} } } }, \label{Eqn:  Non-homogeneous Connection Prob 2}   
\end{eqnarray}
where the last equality follows from the nearest BS distance distribution for $R_i$. 
 Using \eqref{Eqn: Non-homogeneous Connection Prob 2}, we can write $p_k^\star$ as
$
p_k^\star = \int_0^\infty \prod_{ \myatop{i=1}{i \neq k}  }^K \exp\paren{-{ \Lambda_i\paren{ \mathcal{B}\paren{ \vecbold{ X^{(0)} }, Q_i^{(k)}(u)} } } } f_{R_k}(u) du.  \label{Eqn:  Non-homogeneous Connection Prob 3}
$

\section{The Proof of Lemma \ref{Lemma: Association Probability} } \label{Appendix: Association Probability}
For the case of spatial homogeneous PPPs, we just need to replace ${ \Lambda_i\paren{ \mathcal{B}\paren{ \vecbold{ X^{(0)} }, Q_i^{(k)}(u)} } }$ with $\pi \lambda_i \paren{Q_i^{(k)}(u)}^2$. Without loss of generality, we assume that the location of the test user is at the origin, i.e., $\vecbold{ X^{(0)} } = (0,0)$. Thus, using Lemma \ref{Lemma: Association Probability for non-homogeneous PPPs}, we can write $p_k^\star $ as
\begin{eqnarray}
p_k^\star = \int_0^\infty \prod_{ \myatop{i=1}{i \neq k}  }^K \exp\paren{-\pi \lambda_i \paren{Q_i^{(k)}(u)}^2} f_{R_k}(u) du.  \label{Eqn: Connection Prob 3}
\end{eqnarray}

We note that exponent in \eqref{Eqn: Connection Prob 3} can be written as a summation $\sum_{i=1, i \neq k}^K \lambda_i \paren{Q_i^{(k)}(u)}^2$, and some terms inside the summation may not be active for some particular values of $u$ if $G_k(u) \geq \frac{\beta_i P_i}{\beta_k P_k} G_i(0)$. Recalling the definition of $a_i \defeq \frac{\beta_i P_i}{\beta_k P_k} G_i(0)$, we observe that the condition $G_k(u) \geq \frac{\beta_i P_i}{\beta_k P_k} G_i(0)$ holds if and only if $u \leq G_k^{-1}\paren{a_i}$. Introducing $a_0 = 0$ and $a_{K+1} = +\infty$ to have the integration limits from $0$ to $\infty$, and enumerating $a_i$'s in descending order for $i \neq k$, we finally arrive the desired result 
$
p_k^\star = \sum\limits_{j = 1}^K {\int_{{r_{j - 1}}}^{{r_j}} {\exp \left( { - \pi \sum\limits_{i = 1}^{j - 1} {{\lambda _{\pi \left( i \right)}}{{\left( {Q_{\pi \left( i \right)}^{\left( k \right)}\left( u \right)} \right)}^2}} } \right){f_{{R_k}}}\left( u \right)du} }, \nonumber
$
where $\pi(i)$ is an enumeration of $a_i$'s in descending order, i.e., $a_{\pi(i)} \geq a_{\pi(i+1)}$ for $i=0, \ldots, K-1$ and $r_i = G_k^{-1}\paren{a_{\pi(i)}}$ for $i=0, \ldots, K$.  

\vspace{-0.0cm}
\section{The Proof of Lemma \ref{Non-homogeneous_Conditional_Connection_Distance}} \label{proof_Non-homogeneous_Conditional_Connection_Distance}

In this appendix, we will derive the conditional PDF of the connection distance $R^\star$ given the event $\brparen{A^\star = k}$.  To this end, we will first calculate the conditional CDF of  $R^\star$ given $\brparen{A^\star = k}$, which will be denoted by $F_{R^\star | \brparen{A^\star = k}}(r)$. Let $R_i$ be the nearest distance from $\Phi_i$ to the test user for $i=1, \ldots, K$.  Then,  
\begin{eqnarray}
F_{R^\star | \brparen{A^\star = k}}(r) 
&=& \frac{1}{p_k^\star}\PR{R^\star \leq r \mbox{ and } A^\star = k} \nonumber \\
&=& \frac{1}{p_k^\star} \int_0^r \PR{\bigcap_{ \myatop{i=1}{i \neq k} }^K \brparen{\beta_i P_i G_i(R_i) \leq \beta_k P_k G_k\paren{r}} \bigg| R_k = u} f_{R_k}(u) du.  \label{Eqn: Non-homogeneous Connection Distance 1}
\end{eqnarray}

Using the conditional independence of the events $\brparen{\beta_i P_i G_i\paren{R_i} \leq \beta_k P_k G_k\paren{R_k}}$ for $i \in \brparen{1, \ldots, K} \backslash \brparen{k}$ for any given particular realization of $R_k$ and the independence of the nearest BS distances from different tiers, we can further simplify \eqref{Eqn: Non-homogeneous Connection Distance 1} as
\begin{eqnarray}
F_{R^\star | \brparen{A^\star = k}}(r) &=& \frac{1}{p_k^\star} \int_0^r \PR{\bigcap_{ \myatop{i=1}{i \neq k} }^K \brparen{\beta_i P_i G_i(R_i) \leq \beta_k P_k G_k\paren{u}} \bigg| R_k = u} f_{R_k}(u) du \nonumber \\
&=& \frac{1}{p_k^\star} \int_0^r \prod_{ \myatop{i=1}{i \neq k} }^K \exp\paren{-{ \Lambda_i\paren{ \mathcal{B}\paren{ \vecbold{x}^{(o)}, Q_i^{(k)}(u)} } } } f_{R_k}(u) du. \label{Eqn: Non-homogeneous Connection Distance 2}
\end{eqnarray}

We obtain the conditional PDF of $R^\star$ given $A^\star = k$ by differentiating \eqref{Eqn: Non-homogeneous Connection Distance 2} with respect to $r$. This leads to 
$$
{f_k}\paren{ r } = \frac{1}{{p_k^ \star }} \prod_{ \myatop{i=1}{i \neq k} }^K \exp\paren{-{ \Lambda_i\paren{ \mathcal{B}\paren{ \vecbold{x}^{(o)}, Q_i^{(k)}(r)} } } } f_{R_k}(r),  \label{Eqn: Non-homogeneous Connection Distance 3}
$$
where ${f_{{R_k}}}\left( r \right) = {e^{ - \Lambda_k\paren{ \mathcal{B}\paren{ \vecbold{x}^{(o)}, r } } } }\frac{d}{{dr}} \Lambda_k\paren{ \mathcal{B}\paren{ \vecbold{x}^{(o)}, r } }$ is the nearest tier-$k$ BS distance distribution.

\section{The Proof of Lemma \ref{Conditional_Connection_Distance}} \label{proof_Conditional_Connection_Distance}
Similar to Appendix \ref{Appendix: Association Probability}, the mean measure $\Lambda_i\paren{ \mathcal{B}\paren{ \vecbold{x}^{(o)}, Q_i^{(k)}(r) } }$ reduces to $\pi\lambda_i\paren{Q_i^{(k)}(r)}^2$ when only homogeneous PPPs are considered. Without loss of generality, the location of the test user is assumed to be at the origin, i.e., $ \vecbold{x}^{(o)} = (0,0)$. Thus, we can rewrite ${f_k}\paren{ r }$ as 
\begin{eqnarray}
f_k(r) = \frac{1}{{{p_k^\star}}}\exp \left( { - \pi \sum_{ \myatop{i=1}{i \neq k} }^K  {{\lambda _i}{{\left( {Q_i^{\left( k \right)}\left( r \right)} \right)}^2}} } \right){f_{{R_k}}}\left( r \right) \hspace{0.25cm} \mbox{ for } r \geq 0. \label{Eqn: Connection Distance 3}
\end{eqnarray}
We observe that the summation term appearing in \eqref{Eqn: Connection Distance 3} is exactly the same one appeared in \eqref{Eqn: Connection Prob 3}. Hence, the same enumeration step can be carried out to arrive at the final result 
\begin{eqnarray}
f_k(r) = \frac{1}{{{p_k^\star}}}\sum\limits_{j = 1}^K {\exp \left( { - \pi \sum\limits_{i = 1}^{j - 1} {{\lambda _{\pi \left( i \right)}}{{\left( {Q_{\pi \left( i \right)}^{\left( k \right)}\left( r \right)} \right)}^2}} } \right){f_{{R_k}}}\left( r \right)} {1_{\left\{ { r \in \left[ {{r_{j - 1}},{r_j}} \right)} \right\}}}, \nonumber
\end{eqnarray}
where $a_0 = 0$, $a_{K+1} = +\infty$, $a_i = \frac{\beta_i P_i}{\beta_k P_k} G_i(0)$ for $i \in \brparen{1, \ldots, K} \backslash \brparen{k}$, $\pi(i)$ is an enumeration of $a_i$'s in descending order, i.e., $a_{\pi(i)} \geq a_{\pi(i+1)}$ for $i=0, \ldots, K-1$ and $r_i = G_k^{-1}\paren{a_{\pi(i)}}$ for $i=0, \ldots, K$.


\ifCLASSOPTIONcaptionsoff
  \newpage
\fi



%

%

\vfill \vfill
\end{document}